\documentclass[prc,showpacs,preprint]{revtex4}

\usepackage{amssymb}
\usepackage{graphicx}
\usepackage{dcolumn}
\usepackage{bm}
\usepackage{multirow}
\usepackage{color}

\begin{document}

\title{Effects of thermal shape fluctuations and pairing fluctuations on the giant dipole resonance in warm nuclei}
\author{A. K. Rhine Kumar$^1$}
\email{rhinekumar@gmail.com}
\author{P. Arumugam$^1$}
\author{N. Dinh Dang$^{2,3}$}
\affiliation{$^1$Department of Physics, Indian Institute of Technology Roorkee,
Uttarakhand - 247 667, India\\
$^2$Theoretical Nuclear Physics Laboratory, Nishina Center for Accelerator-Based Science, RIKEN 2-1 Hirosawa, Wako City, 351-0198 Saitama, Japan,
 $^3$Institute for Nuclear Science and Technique, Hanoi, Vietnam}
\date{\today}

\begin{abstract}
Apart from the higher limits of isospin and temperature, the properties of atomic nuclei are intriguing and less explored at the limits of lowest but finite temperatures. At very low temperatures there is a strong interplay between the shell (quantal fluctuations), statistical (thermal fluctuations), and residual pairing effects as evidenced from the studies on giant dipole resonance (GDR). In our recent work [Phys. Rev. C \textbf{90}, 044308 (2014)], we have outlined some of our results from a theoretical approach for such warm nuclei where all these effects are incorporated along within the thermal shape fluctuation model (TSFM) extended to include the fluctuations in the pairing field. In this article, we present the complete formalism based on the microscopic-macroscopic approach for determining the deformation energies and a macroscopic approach which links the deformation to GDR observables. We discuss our results for the nuclei $^{97}$Tc, $^{120}$Sn, $^{179}$Au, and $^{208}$Pb, and corroborate with the experimental data available. The TSFM could explain the data successfully at low temperature only with a proper treatment of pairing and its fluctuations. More measurements with better precision could yield rich information about several phase transitions that can happen in warm nuclei.
\end{abstract}

\maketitle
\section{Introduction}
The study of giant dipole resonance (GDR)  at high temperature ($T$) and angular
momentum ($J$) has been an interesting area of research which has revealed
several structural properties of nuclei at extreme conditions. Being a fundamental
mode of photo excitation, GDR can probe nuclei at extreme conditions and
even those with exotic structures \cite{PRL113}. Several earlier
works on GDR focused on the high-$J$ regime \cite{Snov86,Gaar92} whereas
 recent studies at extreme isospins have potential astrophysical implications
\cite{PRL111rossi,HaraPRL105,HaraPRC81,PRC044301}. The GDR at  low $T$ is also relatively less explored and its studies have gained acceleration in recent times \cite{Dang:531,mukhu9,Pandit2012434,Balaram}. Experimentally, it is very difficult to populate  nuclei at low excitation energies, but
it is still feasible due to recent developments in the experimental facilities. Several properties of nuclei at low $T$ are still not clear: for example, the existence of pairing phase transition, the order of it if it exists, the role of fluctuations, etc. The low-$T$ region is quite intriguing because
the microscopic effects such as shell (quantal) and pairing effects are quite active, although they are modified by thermal effects. Since the nucleus is a tiny system, the thermal fluctuations inherent in finite systems are expected to be large. The shape degrees of freedom being crucial for  nuclear
structure, the deformation parameters are closely associated with the order parameters for the related transitions. Hence the thermal shape fluctuations (TSF) are the most dominant ones, and at low $T$ the fluctuations in the pairing field can also contribute significantly. Many models \cite{ALHAT,aruprc1,MORET40B,Dang064315,Dang044303}
have been used to study  the effect of both these fluctuations separately, but the combined effect of these two was not investigated until our
recent efforts \cite{rhineaip1,Rhineprc}. 

Various theoretical approaches have been introduced to investigate the   GDR. In a macroscopic approach, GDR is a collective  mode of excitation of nuclei caused by the out-of-phase oscillations between the proton and neutron fluids under the influence of the electromagnetic field induced by an emitted/absorbed photon. Here, the GDR couples directly to the shape of the nucleus hence providing corresponding structure information. The thermal shape fluctuation model (TSFM) is based on such a macroscopic approach
and explains the increase in GDR width with $T$ by taking in to account the average of GDR cross sections over the shape degrees of freedom. The TSFM is quite successful in interpreting several GDR measurements at high $T$ and $J$ and hence is often used by experimentalists as theoretical support
for their data. 

In an alternative  approach, the GDR observables can be calculated utilizing
a linear response theory incorporating the TSF within the static path approximation \cite{ansari011302,ansari024310}. Other rigorous microscopic approaches exist \cite{PRL109,PRL111} but are mostly limited to the study of low-lying states.
In microscopic approaches, such as the phonon-damping model (PDM), GDR damping is explained through coupling of the
GDR phonon to particle-hole, particle-particle, and hole-hole excitations \cite{DangPLB,Dang044303,Dang:636,Dang:645,Dang044333,Dang:531}. The PDM  can explain the increase in GDR width with $T$, at moderate $T$ and the width saturation at high $T$.

Apart from the above discussed models, a few phenomenological parametrizations have been reported \cite{Kusn98,Pandit2012434}, which are very successful in explaining the global trend of the GDR width as a function of $T$ and $J$. These parametrizations are constructed to mimic the results of TSFM
and hence have only empirical basis without any microscopic or macroscopic
foundation. However, in recent literature \cite{pandit054327,pandit044325,Pandit2012434,Balaram}, these parametrizations are referred synonymously to models and hence should not be understood as a variant of TSFM.  

The damping of GDR width at low $T$ and the strong influence of pairing
on GDR width were first suggested by one of the present authors \cite{Dang:531}. The first low-$T$ GDR width measurements were reported in Ref.~\cite{Heck43} by measuring the $\gamma$ decays in coincidence with $^{17}$O particles scattered inelastically from $^{120}$Sn. These  data were successfully explained by using  the quasiparticle representation of the PDM \cite{Dang044303} incorporating  the thermal pairing correlations. These calculations are based on the modified
BCS approach, where the pairing gap does not vanish abruptly, but decreases slowly with increasing $T$ \cite{Dang014318,Dang:064319,Dang:014304,MORET40B}. In the conventional BCS approach, the pairing gap collapses at a critical temperature of the transition from the superfluid phase to the normal one, which is equal to $T_c \simeq 0.57\Delta(0)$, with $\Delta(0)$ being the pairing gap at $T=0$. In another important low-$T$ measurement, the  GDR width in $^{179}$Au \cite{camera155} at $T=0.7$ MeV was observed to be similar to the ground-state value, in contrast to the expectation of a larger width due to the thermal excitations. The shell effects were predicted \cite{camera155} to explain this quenching, like in the case of $^{208}$Pb. However, in Ref.~\cite{aruepl} it was reported that a proper inclusion of shell effects leads to an increase in  GDR width in $^{179}$Au and there is a need for an exact treatment of the pairing fluctuations (along with the TSF) to overcome this discrepancy. Subsequently, the importance of considering pairing in the TSFM has been reported in Refs.~\cite{rhineaip1,aruriken,Dangarx}. 

Very recently,  GDR measurements in the low-$T$ regions were carried out  at the Variable Energy Cyclotron Centre, Kolkata \cite{Balaram,mukhu9,Pandit2012434,pandit044325}
and highlighted the interesting nuclear properties at low $T$.   
In a recent work \cite{Balaram}, it has been mentioned that it would also
be interesting to compare the data with TSFM calculations by including the effect of thermal pairing. In some of these recent works \cite{mukhu9,Pandit2012434,Balaram}, the empirical parametrizations (which mimic the results of TSFM at higher $T$) were reported to fail in explaining the experimental data at low $T$. It has to be noted that, at low $T$, the application of even the proper TSFM is incomplete and hence the corresponding parametrizations  would also fail. This difficulty in such a parametrization was overcome \cite{Pandit2012434} by introducing another empirical entity, i.e., the GDR width at critical $T$ (which corresponds to the width observed at lowest $T$), and the GDR width below that critical $T$ was assumed to be the same constant value \cite{Pandit2012434}. 

In the present work we discuss a proper TSFM that is also applicable at low
$T$.  The success of a modified pairing approach \cite{Dang044303,Dang:064319,Dang:014304,Balaram,Dang044333} at low-$T$ and that of the TSFM elsewhere \cite{aruprc1,ALHA,Kusn98,Orma97,aruepl,aruepja} have motivated us to consider a combination of  pairing correlations within the TSFM. Preliminary results of our approach can be found in Refs.~\cite{rhineaip1,aruriken,Dangarx}.
A short discussion on our approach and some important results for the nuclei
$^{97}$Tc, $^{120}$Sn, and $^{208}$Pb were reported recently \cite{Rhineprc}.
Here we present for the first time our complete formalism in detail along with elaborate discussions of our results for the nuclei $^{97}$Tc, $^{120}$Sn, and $^{208}$Pb. Additionally, we present our results also for the nucleus $^{179}$Au. 
\section{Theoretical Framework}
The theoretical approach we follow is explained in four parts
corresponding to (A) the finite temperature Nilsson-Strutinsky method for
deformation energy calculations, (B) the BCS approach for pairing, (C) the
macroscopic
approach for GDR which relates the shapes to GDR observables, and (D) thermal fluctuations in finite systems.

\subsection{Finite temperature  Nilsson-Strutinsky method}\label{deforma_e}
For deformation energy calculations, we adopt a formalism based on the finite temperature  Nilsson-Strutinsky method \cite{aruprc1}, which is extended to include thermal pairing. Regarding  Strutinsky's (microscopic-macroscopic) prescription \cite{STRU1}, the total free energy ($F_{\mathrm{TOT}}$) at a fixed deformation is calculated by using the expression
\begin{equation}
F_{\mathrm{TOT}}=E_{\mathrm{LDM}}+\sum_{P,N}\delta F\;. 
\label{FTOT}
\end{equation}%
The sum at the right-hand side runs over protons ($P$) and neutrons ($N$). The macroscopic part, i.e., the liquid-drop energy ($E_{\mathrm{LDM}}$), is calculated by summing up the Coulomb and surface energies \cite{ARU1,RAMS} corresponding to a triaxially deformed shape defined by the deformation parameters $\beta $ and $\gamma $ (\ref{Hill wheeler}) as
\begin{equation}
E_{\mathrm{LDM}}(\beta,\gamma)=\left\{\left[B_s(\beta,\gamma)-1\right]+2\chi\left[B_c(\beta,\gamma)-1\right]a_s\right\}A^{2/3}.
\label{Eldm}
\end{equation}
The function $B_s(\beta,\gamma)$ gives the dependence of the surface energy
on shape and is equal to the dimensionless ratio of the surface area of the shape in question to the area of the original sphere. The function $B_c(\beta,\gamma)$ is the dimensionless ratio of the electrostatic energy of a distorted shape distribution to that of the sphere. Both $B_s(\beta,\gamma)$ and $B_c(\beta,\gamma)$ are elliptic integrals which are evaluated for a given deformation. The parameters $a_{s}$ and $\chi$ are chosen to be $19.7$ MeV and  $\frac{Z^2}{45A}$, respectively,
where $Z$ and $A$ are the charge and mass numbers of the nucleus. It has to be noted that $E_\mathrm{LDM}(\beta,\gamma)$ is the binding energy relative to that of a sphere and hence is termed the deformation energy. Such a quantity is enough for us as we study only the variation of energy over the deformation space rather than its absolute value. At finite $T$, in principle,  one should consider that the free energy and the liquid drop parameters can have a $T$ dependence. However, any effect of $T$ on the liquid drop energy is found to be negligible at lower  $T\ (\lesssim 2$ MeV) \cite{Guet205}.
This will be further attenuated in the deformation (relative) energy that
we are interested in. At  higher $T$, the shell corrections vanish and hence the equilibrium shape becomes spherical where the deformation energy is zero.
  
The calculation of the microscopic part of the free energy (\ref{FTOT}) is based on  the Nilsson Hamiltonian given by \cite{NilRag,Chandra333},
\begin{equation}
H_{0}=\sum_{i}\frac{\overrightarrow p^2}{2m}+\frac{1}{2}m\sum_{k=1}^{3}\omega_{k}^{2}x_{k}^{2}+C
\overrightarrow l.\overrightarrow s+D(\overrightarrow l^2-2\langle\overrightarrow l^2\rangle)\;,  \label{Nilsson_H}
\end{equation}
where the index $i$ represents the sum over all single particles. The three oscillator frequencies $\omega_k$, where $k=1, 2, 3$, are given by the Hill-Wheeler approximation \cite{Hill89} as
\begin{equation}
\omega_k=\omega_{0}\exp\left[-\sqrt \frac{5}{4\pi}\beta\cos\left(\gamma-\frac{2\pi
k}{3}\right)\right]\;,
\label{Hill wheeler}
\end{equation}
with the constraint of constant volume for equipotentials: $\omega_x\omega_y\omega_z=\mathring\omega_{0}^{3}$=
constant. The oscillator frequency is chosen as, $h\mathring\omega_{0}=\frac{45.3}{A^{1/3}+0.77}$
MeV and the Nilsson parameters $\kappa$ and $\mu$ values are chosen for appropriate
mass regions from Refs.~\cite{Anderson309,Chandra333,Nilsson}.  The Hamiltonian (\ref{Nilsson_H}) is diagonalized in cylindrical representation up to first
twelve principal harmonic oscillator shells using the matrix elements given in \cite{NUCMOD} to obtain the singleparticle energies ($e_{i}$), separately for protons and neutrons.

While considering the  pairing fluctuations (PF), we use the grand canonical ensemble (GCE), where the particle number fluctuations are allowed \cite{MORET40B,MORET} by fixing the chemical potential ($\lambda$). The free energy  corresponding to the thermal average within the GCE is
\begin{equation}\label{free energy}
F=\left\langle H_{0}\right\rangle -\lambda N_p-TS\;,
\end{equation}%
where $H_{0}$ is the nuclear Hamiltonian (3) which is independent of
$T$, $N_p$ is the particle number, and $S$ is the entropy. The above expression can be rewritten as%
\begin{equation}
F=\sum_{i}(e_{i}-\lambda -E_{i})-2T\sum_{i}\ln [1+\exp (-E_{i}/T)]+\frac{%
\Delta ^{2}}{G}\;,
\end{equation}
where $E_{i}=\sqrt{(e_{i}-\lambda )^{2}+\Delta ^{2}}$ are the
quasiparticle energies. $\Delta$ is the pairing gap, $G$ is the pairing
strength and they are determined using the procedures mentioned in Sec.~\ref{pairing}. It has to be noted that even in self-consistent Hartree-Fock
calculations \cite{BQ81}, for a fixed deformation, the $T$ dependence of
$e_i$ is negligible.

In the Strutinsky way, the smoothed free energy can be written as \cite{BQ81}
\begin{eqnarray}
\widetilde{F}&=&2\sum_{i}(e_{i}-\lambda )\widetilde{n}_{i}-2T\sum_{i}\widetilde{s%
}_{i}{-\sum_i {\Delta_i}\widetilde{k_i}}\nonumber\\
&&+2\gamma _{s}\int_{-\infty }^{\infty }\widetilde{f}(x)x%
\sum_{i}n_{i}(x)dx
\label{F_tilde}
\end{eqnarray}
with the last term included to give better plateaus conditions \cite{aruprc1}.
Here $\widetilde{f}(x)$ is the averaging function given by%
\begin{equation}
\widetilde{f}(x)=\frac{1}{\sqrt{\pi }}\exp
(-x^{2})\sum_{m=0}^{q}C_{m}H_{m}(x)\;; 
\label{FTIL}
\end{equation}%
$C_{m}=(-1)^{m/2}/[2^{m}(m/2)!]$ if $m$ is even and $C_{m}=0$ if $m$ is odd;
$x=(e-e_{i})/\gamma _{s}$, $\gamma _{s}$ is the smearing parameter
satisfying the plateau condition $d\widetilde{F}/d\gamma _{s}=0$; $\;q$ is
the order of smearing and $H_{m}(x)$ are the Hermite polynomials. The
averaged occupation numbers,  single-particle entropies, and pairing numbers are given by
\begin{equation}
\widetilde{n}_{i}=\int_{-\infty }^{\infty }\widetilde{f}(x)\;n_{i}(x)\;dx\;,
\label{NITT2}
\end{equation}%
\begin{equation}
\widetilde{s}_{i}=\int_{ -\infty }^{\infty }\widetilde{f}(x)\;s_{i}(x)\;dx\;,
\label{SITT2}
\end{equation}%
\begin{equation}
\widetilde{k}_{i}=\int_{ -\infty }^{\infty }\widetilde{f}(x)\;k_{i}(x)\;dx\;,
\quad\text{}\label{KITT2}
\end{equation}%
respectively. These integrations and the one in Eq.~(\ref{F_tilde}) are carried out numerically. In such a case there could be numerical uncertainties at
very low $T\ (\lesssim 0.2$ MeV). To avoid such problems it would
be nice to  extend the maximum term approximation  method \cite{CIVI82,aruprc1} to include pairing. The quasiparticle occupation numbers resulting from both thermal and pairing effects are given by%
\begin{equation}
n_{i}=\frac{1}{2}\left[ 1-\frac{e_{i}-\lambda }{E_{i}}\tanh \left( \frac{%
E_{i}}{2T}\right) \right]
\end{equation}
and the quasiparticle occupation numbers owing to thermal effects alone are given by%
\begin{equation}\label{nit}
n_{i}^{T}=\frac{1}{1+\exp (E_{i}/T)}
\end{equation}
so that the total entropy can be written as%
\begin{equation}
S=2\sum_{i}s_{i}=-2\sum_{i}\left[ n_{i}^{T}\ln n_{i}^{T}+(1-n_{i}^{T})\ln
(1-n_{i}^{T})\right]\;.
\end{equation}
The pairing numbers are given by
\begin{equation}
k_i=\frac{\Delta}{2E_i}\tanh\left(\frac{E_i}{2T}\right).
\end{equation}
Calculating the smooth part  of $\frac{\Delta ^{2}}{G}$ in this way, we come
across unrealistic results such as  a large proton pairing gap for the closed-shell nucleus $^{120}$Sn,  and a subsequent well deformed ($\beta\sim0.2$) equilibrium shape at $T=0.1$ MeV. Inherently, the Strutinsky method to incorporate pairing leads to an overestimation of pairing gap as discussed in Sec.~\ref{results}.  Also, such a method leads to inconveniences while considering pairing fluctuations [Eq.~(\ref{ave_all})] where the calculations have to be carried out for a given pairing gap. Hence, we replace the third term at the right-hand side
of Eq.~(\ref{F_tilde}) with -$\frac{\Delta ^{2}}{G}$ to obtain the following
expression for the smoothed free energy: 
\begin{eqnarray}
\widetilde{F}&=&2\sum_{i}(e_{i}-\lambda )\widetilde{n}_{i}-2T\sum_{i}\widetilde{s%
}_{i} {- \frac{\Delta ^{2}}{G}}\nonumber\\
&&+2\gamma _{s}\int_{-\infty }^{\infty }\widetilde{f}(x)x%
\sum_{i}n_{i}(x)dx\;.
\label{ftile_f}
\end{eqnarray}
In Eq.~(\ref{ftile_f}) the term $\frac{\Delta^{2}}{G}$ is just a residual correction and the dominant contribution is through the quasiparticle energies,  occupation numbers, and entropies, which are taken care of exactly.   

To quantify the role of pairing, we also carry out the calculations without pairing ($\Delta=0$) by considering the canonical ensemble (CE). In this case,  the free energy is given by \cite{aruprc1}
\begin{equation}\label{free energy_CE}
F=\left\langle H_{0}\right\rangle -TS
\end{equation}
and the corresponding discrete and smoothed energies can be written as
\begin{equation}
F=2\sum_{i}e_{i}n^T_{i}-2T\sum_{i}s_{i}
\end{equation}
where
\begin{equation}
n_{i}^{T}=\frac{1}{1+\exp [(e_{i}-\lambda)/T)]}\;.
\end{equation}
The Strutinsky smoothed free energy is given by
\begin{eqnarray}
\widetilde{F}&=&2\sum_{i}e_{i}\widetilde{n}^T_{i}-2T\sum_{i}\widetilde{s
}_{i}\nonumber\\&&+2\gamma _{s}\int_{-\infty }^{\infty }\widetilde{f}(x)x%
\sum_{i}n^T_{i}(x)dx\;.
\end{eqnarray}
Here we have 
\begin{equation}
\widetilde{n}_{i}^T=\int_{-\infty }^{\infty }\widetilde{f}(x)\;n_{i}^T(x)\;dx\;
\end{equation}%
and all the other quantities are the same as given in the previous section.

\subsection{BCS approach for pairing}\label{pairing}

Here we discuss briefly how we utilize the pairing prescriptions in our
formalism following the BCS approach \cite{BCS57}. 
At finite temperature, the superfluid phase can also exist and in some cases the reentrance of pairing can occur at excited states at nonzero angular momentum. These features  have been well discussed in Refs.~\cite{MORET,Dang054324re,prl105pair}. The BCS equations at finite temperature can be written as   
\begin{eqnarray}
\label{BCST_delta}
\Delta &=& G\sum_iu_iv_i(1-2n_{i}^{T})\;, \\ 
\label{BCST_N}
N_{p} &=& 2\sum_i\{ n_{i}^{T}+(1-2n_{i}^{T})v_i^2\},   
\end{eqnarray}
where
\begin{equation}
\label{VK2}
v_i^2=\frac 12\left[ 1-\frac{(e_i-\lambda )}{E_i}\right] 
\end{equation}
represents the BCS occupation number and $v_i^2+u_i^2=1$.  Pairing can be dealt with either the constant gap approximation or the constant strength ($G$) approximation. In the former approach, $\Delta$ is chosen to be a constant value usually estimated from the odd-even mass differences \cite{BhoMot} or taken as the average empirical value $12/\sqrt{A}$ MeV. In such a case only Eq.~(\ref{BCST_delta}) has to be solved to obtain $\lambda$ at $T=0$.
For calculations without pairing fluctuations, we follow the constant-$G$
approach with the value of $G$ evaluated at $T=0$ and solve Eqs.~(\ref{BCST_delta}) and (\ref{BCST_N}) simultaneously. When we consider the pairing fluctuations, we solve only Eq.~(\ref{BCST_N}) for a given $\Delta$ which is a variable of integration [Eq.~(\ref{ave_all})]. The value of $G$ can be obtained from  either Strutinsky calculations \cite{FUNNY} or by fitting with empirical data \cite{Nilsson} as outlined in the next section.

Our excited state can have pairs in thermal equilibrium, which replaces the
ground-state ($T=0$) excited pairs and  single particles in excited states. The pairs in thermal equilibrium follow the distribution defined by $v_i^2$, if we neglect the influence of excited pairs and excited single particles. The excited pairs and excited single particles follow the distribution defined by $n_{i}^{T}$~(\ref{nit}). Hence we can assume that the resultant distribution is the combined effect of these two distributions.

\subsubsection{$G$ from Strutinsky calculations}\label{p_constant1}
In this method \cite{BOLST}, the pairing gap corresponding to a smooth (Strutinsky
smeared) distribution of single-particle states is assumed to be the same value obtained from the empirical average pairing gap, i.e., $\widetilde\Delta=\frac{12}{\sqrt A}$ MeV. Then the pairing force strength can be calculated using the expression \begin{equation}
\frac{1}{G}=\widetilde{\rho}\ln\left\{\left[\left(\frac{N_{p}}{2\widetilde{\rho}\widetilde{\Delta}}\right)^2+1\right]^{(1/2)}\frac{N_{p}}{2\widetilde{\rho}\widetilde{\Delta}}\right\},
\end{equation}
where
\begin{equation}
\widetilde{\rho}=\frac{1}{2}\widetilde{g}(\widetilde\lambda)\;
\end{equation}
is the average density of pairs at the Fermi surface.
 The average level density at the Fermi level [$\widetilde{g}(\widetilde\lambda)$] can be calculated using the relation 
\begin{equation}\label{average LD}
\widetilde{g}(e)=\frac{1}{\gamma \sqrt{\pi }}\sum_i \exp(-x^{2})\sum_{m=0}^{q}C_{m}H_{m}(x)\;.
\end{equation}
The  Fermi energy ($\widetilde\lambda$) can be calculated by solving the equation $\widetilde{n}(\widetilde\lambda)=N_{p}$, and the average particle number (considering the states from bottom to a given energy $e$ to be filled) can be evaluated using the relation
\begin{equation}
\widetilde{n}(e)=\sum_{i}\left\{ \frac{1}{2}[1+\mathrm{erf}(x)]-\ \frac{1}{\sqrt{\pi }}\exp(-x^{2})\sum_{m=1}^{q}C_{m}H_{m-1}(x)\right\}\;. 
\end{equation}
In these calculations, the single-particle states comprise all the states below the Fermi level and an equal number of states above it. 

\subsubsection{$G$ from empirical data \label{pconstant_2}}
In an alternate approach, the  pairing strength constant can be calculated from an empirical fit to odd-even mass differences and their $1/\sqrt A$ dependence. Such a fit has been carried out in Ref.~\cite{Nilsson} for the modified oscillator potential, yielding the values.
\begin{equation}
G_{P,N} =  [19.2 \pm 7.4(N-Z)]/A^2 .
\end{equation}
In these calculations $\sqrt{15Z}$ and $\sqrt{15N}$ single-particle states above and below the Fermi level are included. It has to be noted that the results depend on this choice of configuration space.

\subsection{Macroscopic approach for GDR }
We  follow  a macroscopic approach to relate the nuclear shapes with the GDR observables \cite{aruprc1,THIA1,THIA2}. In this formalism the GDR Hamiltonian
could be written as
\begin{equation}
H=H_{osc}+H_{int}\ ,\label{GDR hamiltonian1}
\end{equation}
where $H_{osc}$ stands for the anisotropic harmonic oscillator Hamiltonian
and $H_{int}$ characterizes the separable dipole-dipole interaction given by
\begin{equation}
H_{int}=\eta\sum_{k=x,y,z}\frac{m\omega_k^2}{2A}\left[\sum_{\nu=1}^A\tau_{3}^{(\nu)}x_k(\nu)\right]^2\;,
\end{equation}
where $\tau_{3}^{(\nu)}$ is the third projection of the Pauli isospin matrix
\begin{equation}
\tau_3=\left( \begin{array}{cc}
1 & 0 \\
0 & -1 \\
\end{array} \right) 
\end{equation}
and $\eta$ is a parameter that characterizes the isovector component of the
neutron or proton average field
\begin{equation}
V_{(N,P)}(\nu)=\frac{m}{2}\left[1\pm\eta \frac{N-Z}{A}\right]\sum_{k=x,y,z}\omega_{k}^2x_{k}^2(\nu)\;.
\end{equation}
In terms of the dipole operator $D$, the GDR Hamiltonian (\ref{GDR hamiltonian1}) can be rewritten as
\begin{equation}
H=H_{osc}+\eta D^\dagger D\;.
\end{equation}
  Including the pairing
interactions in a simple way, the above equation can be modified as%
\begin{equation}
H=H_{osc}+\eta D^{\dagger}D +\chi \mathcal{P}^{\dagger}\mathcal{P}, 
\label{GDRhamiltonian}
\end{equation}
where $\chi $ and $\mathcal{P}$ denote the strength and operator of the pairing interaction,
respectively. In a simple oscillator description, the only relevant role
of the pairing
interaction is to  change the oscillator frequencies [$\omega _{\nu }^{osc}$
($\nu
=x,y,z$)], resulting in the new set of frequencies%
\begin{equation}\label{gdr_frequency}
\omega _{\nu }=\omega _{\nu }^{osc}-\chi \omega ^{\mathcal{P}}\ ,
\end{equation}
where
\begin{equation}\label{omega_P}
\omega ^{\mathcal{P}}=\left(\frac{Z\Delta _{P}+N\Delta _{N}}{Z+N}\right)^{2} .
\end{equation}
Here $\chi$ having the units of MeV$^{-1}$  has to be determined empirically
and hence the associated sign is just notional with an expectation that the
pairing will dampen the oscillations. Alternatively, the  role of pairing can be conceived as  to renormalize the dipole-dipole interaction strength such that, 
\begin{equation}\label{gdr_eta}
\eta=\eta_0-\chi_0\sqrt T\omega^{\mathcal{P}}, 
\end{equation}
with $\chi_0$ having the units of MeV$^{-5/2}$.
This assumption is based on the fact that the dipole-dipole interaction
in its separable form will change only the oscillator frequencies. The pairing interaction also does the same and  hence it is possible to combine these two effects as in the above relation. If pairing has to be introduced as per Eq.~(\ref{gdr_frequency}) then the parameter $\eta$ has to be renormalized accordingly. The $\omega^\mathcal{P}$ depends on $T$ as inherited from $\Delta_P$ and $\Delta_N$. Hence when pairing vanishes, we have $\eta=\eta_0$, allowing us to retain the parametrization of $\eta$ as in our previous approaches without pairing, and hence a meaningful comparison between these two approaches
is possible. In the case of having nonvanishing $\Delta$ at larger $T$, an asymptotic value, say $\Delta\sim0.75$ MeV ($T=2$ MeV) has to be subtracted from the $\Delta$ appearing in Eq.~(\ref{omega_P}), to retain the asymptotic behavior ($\left.\eta\right |_{T\geq2 \text{ MeV}} = \eta_0$). The presence of $T$ in Eq.~(\ref{gdr_eta}) ensures $\left.\eta\right |_{T=0} = \eta_0$ at any $\Delta$.  The exponent of $T$ is chosen on empirical basis
which can be improved further if more low-$T$ data are available.

Including the dipole-dipole and pairing interactions, the GDR frequencies in the laboratory frame are obtained as
\begin{equation}
\widetilde{\omega }_{z}=(1+\eta )^{1/2}\omega _{z}\;,\label{GDR_wz}
\end{equation}
\begin{eqnarray}
\widetilde{\omega }_{2,3} = && \left\{(1+\eta )\frac{\omega
_{y}^{2}+\omega _{x}^{2}}{2} \right. \nonumber \\ && \left.\pm\frac{1}{2}\left\{ (1+\eta )^{2}(\omega _{y}^{2}-\omega
_{x}^{2})^{2}\right\}^{1/2}\right\}^{1/2}.\label{GDR_wxy}
\end{eqnarray}
The above relations look exactly same as the nonrotating limits of the GDR
frequencies given in Refs.~\cite{THIA1,aruepja,aruprc1,aruepl}, but here $\eta$ is redefined to include the effect of pairing. From these GDR frequencies (energies), the GDR cross sections are constructed as a sum of Lorentzians given by
\begin{equation}
\sigma (E_{\gamma })=\sum_{i}\frac{\sigma _{mi}}{1+\left( E_{\gamma
}^{2}-E_{mi}^{2}\right) ^{2}/E_{\gamma}^{2}\Gamma _{i}^{2}}\;,
\label{Eq.Cross}
\end{equation}
where $E_{m}$, $\sigma _{m}$, and $\Gamma$ are the
resonance energy, peak cross section, and full width at half maximum,
respectively \cite{THIA1,THIA2,HILT}. Here $i$ represents the number of components of the GDR which is $3$ as given by Eqs.~(\ref{GDR_wz}) and (\ref{GDR_wxy}). $\Gamma_{i}$ is assumed to depend on the centroid energy through the relation \cite{ALHwidth}
\begin{equation}
\Gamma_{i}= \Gamma_{0}(E_{i}/E_{0})^{\delta}\;,  \label{PowLaw1}
\end{equation}
where $\Gamma_{0}$ and $E_{0}$ are the  resonance width and energy in the
case of a spherical nucleus. In Ref.~\cite{CARL}, an empirical fit between $\Gamma_{i}$ and $E_{i}$ was carried out for the components of the ground-state GDR in several nuclei, yielding $\delta=1.9\pm0.1$ and $\Gamma_{0}/E_{0}^{\delta}=0.026\pm0.005$.
 Hence, the energy dependence of GDR width can be approximated as 
\begin{equation}
\Gamma_{i}\approx 0.026 E_{i}^{1.9}. \label{PowLaw2}
\end{equation}
The value of $\delta$ can be fine tuned to obtain a better fit with the experimental data. We obtained optimal fits with $\delta=1.8$ for $^{97}$Tc and $\delta=1.9$ for $^{120}$Sn, $^{179}$Au, and $^{208}$Pb nuclei.  The peak cross section $\sigma _{m}$ is given by \cite{aruprc1}
\begin{equation}\label{sum_rule}
\sigma _{m}=60\frac{2}{\pi }\frac{NZ}{A}\frac{1}{\Gamma }\;(1+\alpha )\;,
\end{equation}
where $\Gamma$ is the full width at half maximum of the GDR cross section. It may not be appropriate to generalize Eq.~(\ref{sum_rule}) for individual components. But it will not be possible to fix the three peak cross sections ($\sigma_{m}$) from one constraint, which is the GDR energy-weighted sum rule. Hence, Eq.~(\ref{sum_rule}) just sets the dependency of  $\sigma_{mi}$ on $\Gamma_{i}$ with a proportionality constant $\alpha$ which is given by the sum rule.  The parameter $\alpha $  is fixed at 0.3 for all the nuclei. In most of the cases we normalize the peak with the experimental data and hence the choice of $\alpha $ has negligible effect on the results. The other parameters $\eta_0$ (or $\eta$) and $\chi_0$ (or $\chi$) vary with nuclei so that the experimental ground-state GDR width is reproduced. The choices of the parameters $\eta_0$ and $\chi_0$ for different nuclei are given in Table~\ref{table_eta}. The effects of quantal fluctuations other than those represented by the shell effects, such as the particle-number fluctuations on the pairing gap, etc., are just residual and small. We assume that while they are present in the measured  width of the GDR built on the ground state, they are included in our approach by adjusting the parameters $\eta_0$ and $\chi_0$  to reproduce that measured width (at $T$=0). Weakening of such effects with $T$ is effectively represented through the weakening pairing gap. 

\begin{table}
\begin{tabular}{ccc}
\hline
Nucleus & $\eta_0$  & $\chi_0$(MeV$^{-5/2}$) \\
\hline
$^{97}$Tc&2.60&1.7\\  
$^{120}$Sn&2.60&3.5\\
$^{179}$Au&2.25&3.0\\
$^{208}$Pb&3.40&2.5\\ 
\hline
\end{tabular}
\caption{The  parameters $\eta_0$ and $\chi_0$ used for different
nuclei in our calculations.}
\label{table_eta}
\end{table}

\subsection{Thermal shape and pairing fluctuations}
Since the nucleus is a tiny finite system, thermal
fluctuations related to the appropriate degrees of freedom at a finite excitation
energy are large. While considering the shape degrees of freedom (which are dominant as far as the nuclear structure is concerned), the effective GDR
cross sections depend on the thermal shape fluctuations and they carry information about the probable shape rearrangements~\cite{ALHAT} at finite excitation energy. The general expression for the expectation value of an observable $\mathcal{O}$ incorporating such thermal shape fluctuations has the form~\cite{ALHAO,ALHA} 
\begin{equation}\label{average_1}
\langle \mathcal{O}\rangle _{\beta ,\gamma }=\frac{\int_\beta\int_\gamma \mathcal{D}[\alpha
]\exp\left[{-F_\mathrm{TOT}(T;\beta ,\gamma )/T}\right]\mathcal{O}}{\int_\beta\int_\gamma \mathcal{D}[\alpha
]\exp\left[{-F_\mathrm{TOT}(T;\beta ,\gamma )/T}\right]}\;
\end{equation}%
with the volume element given by $\mathcal{D}[\alpha ]=\beta ^{4}|\sin 3\gamma |\,d\beta \,d\gamma$.

In PF calculations,  we consider a range of pairing gap values, which are closer to the BCS value. The chemical potential ($\lambda$)
is fixed at the BCS value, thus we are allowing the system to have
fewer bound and strong bound pairs. Hence there is a  finite  probability to have pairing even at high $T$ also.

The pairing field is a complex quantity since it also contains a phase  and its appropriate metric for integration is $d|\Delta|^2$. Since the free energy depends only on $\Delta$, the phase of $\Delta$ can be integrated out leading to a metric of the form $\Delta\ d\Delta$ \cite{PRL61}.
By including the pairing fluctuations, we have
\begin{equation}
\langle \mathcal{O}\rangle _{\beta ,\gamma ,\Delta _{P},\Delta _{N}}=\frac{%
\int_\beta\int_\gamma\int_{\Delta_{P}}\int_{\Delta_{N}} \mathcal{D}[\alpha ]\exp \left[{-F_\mathrm{TOT}(T;\beta ,\gamma ,\Delta _{P},\Delta _{N})/T}\right]%
\mathcal{O}}{\int_\beta\int_\gamma\int_{\Delta_{P}}\int_{\Delta_{N}} \mathcal{D}[\alpha ]\exp\left[{-F_\mathrm{TOT}(T;\beta ,\gamma ,\Delta_{P},\Delta _{N})/T}\right]}\;
\label{ave_all}
\end{equation}%
with a volume element given by $\mathcal{D}[\alpha ]=\beta ^{4}|\sin 3\gamma |\,d\beta \,d\gamma \,\Delta_{P}
\,\Delta_{N}\ d\Delta _{P}\ d\Delta _{N}$.

We perform the  TSF calculations  by evaluating numerically the  
integrals in Eq.~(\ref{ave_all}) with the free energy and the observables calculated at every mesh point (the four-dimensional space spanned by deformations and pairing gaps), utilizing the microscopic-macroscopic approach outlined in Sec.~\ref{deforma_e}.

\section{Results and discussion}\label{results}
\subsection{$^{120}$Sn}
\begin{figure}
\includegraphics[width=0.55\columnwidth]{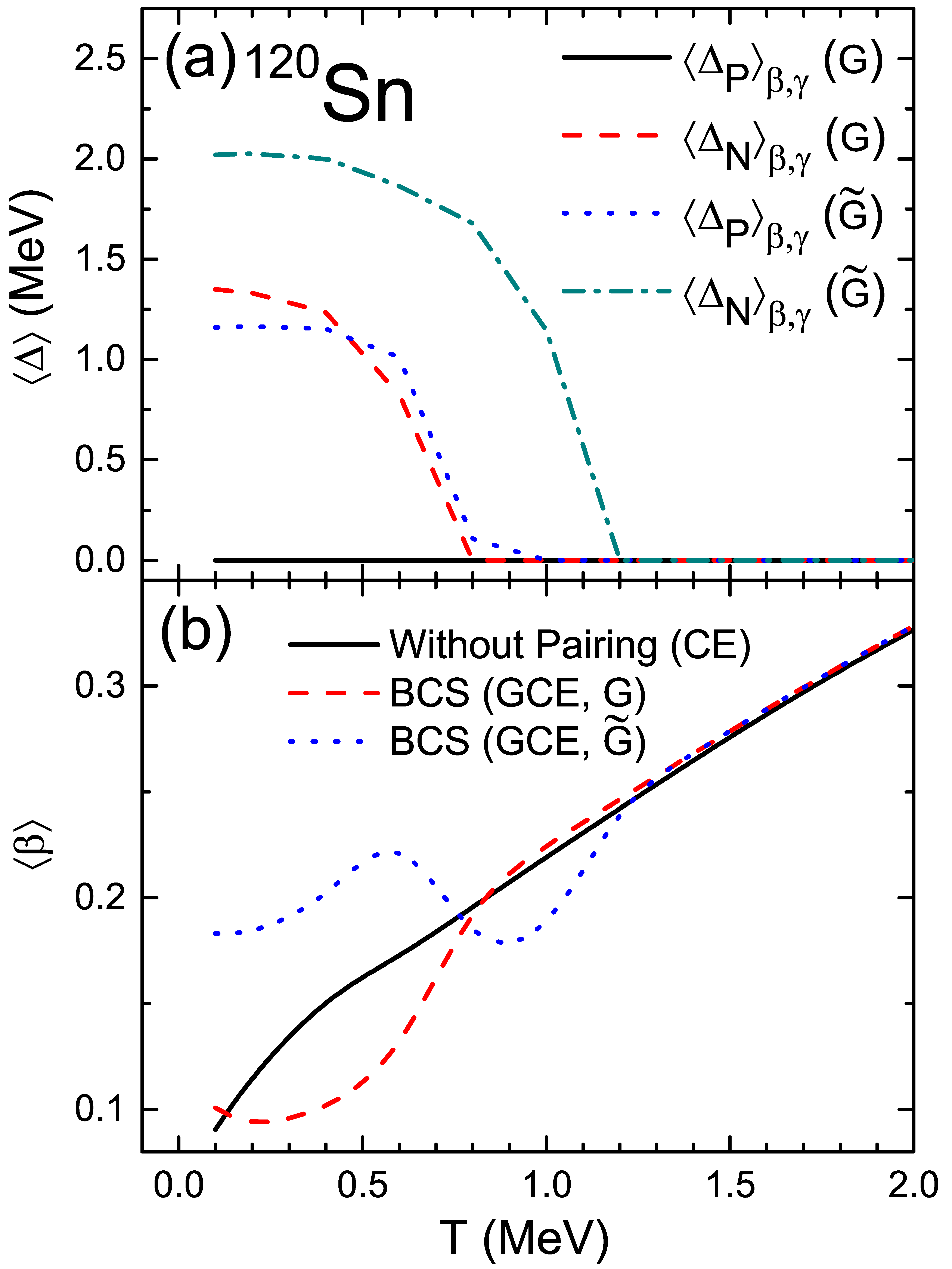}
\caption{(Color online) Role of pairing (within the simple BCS approach) in calculations with thermal shape fluctuations in the case of $^{120}$Sn. (a) Average pairing gap and (b) average quadrupole deformation parameter, as a function of temperature calculated without pairing in  a canonical ensemble approach (CE) and in a grand canonical ensemble approach (GCE) using the pairing strengths obtained from Strutinsky calculations ($\tilde G$) and those quoted in Ref.~\cite{Nilsson}
($G$).}
\label{fig_largeG}
\end{figure}

\begin{figure}
\includegraphics[width=0.55\columnwidth]{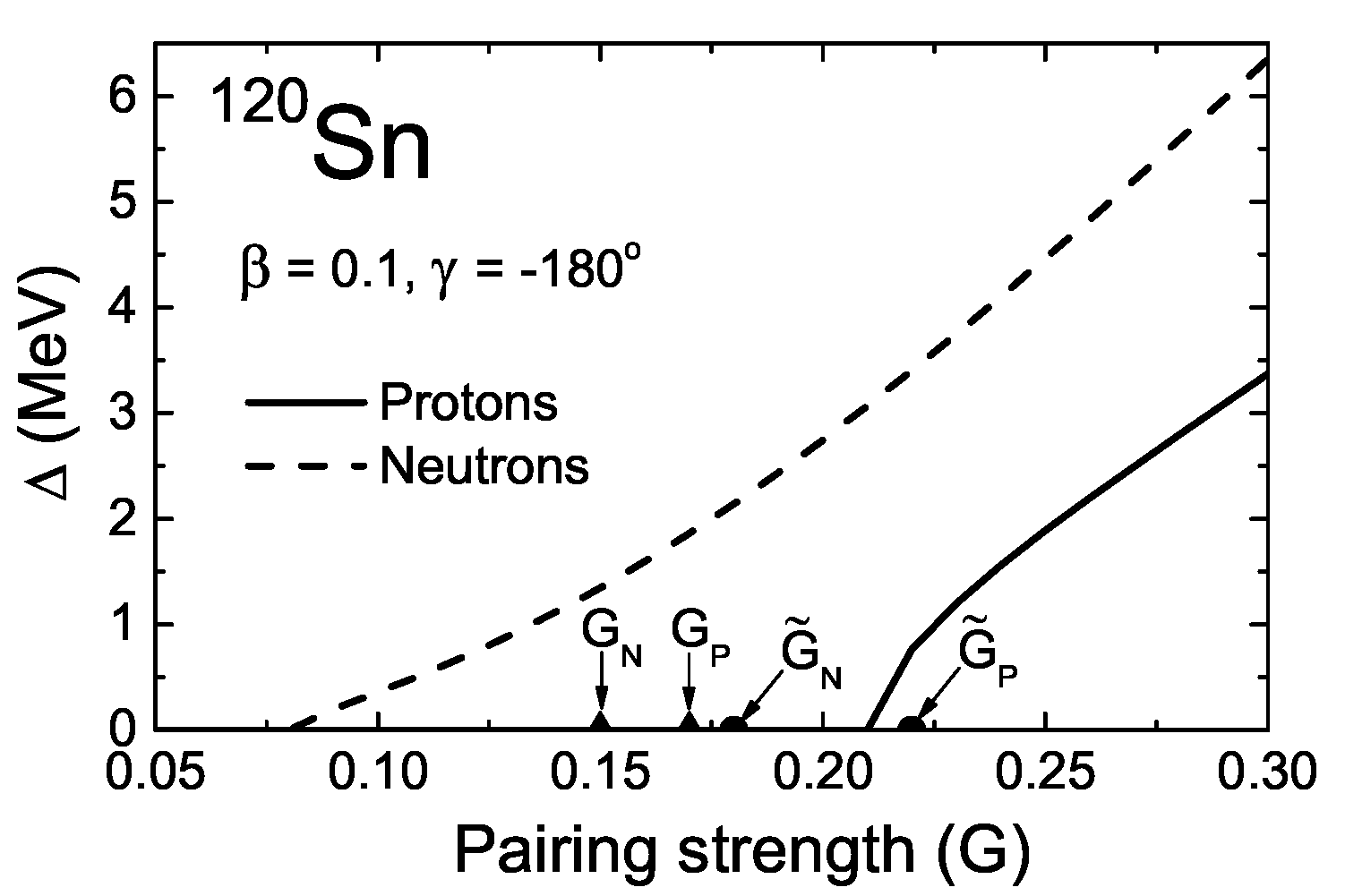}
\caption{The variation of pairing gap for protons (solid line) and neutrons
(dashed line), as a function of pairing strength in
the case of $^{120}$Sn. The prescriptions by Strutinsky calculations ($\tilde G$) and those quoted in Ref.~\cite{Nilsson} ($G$) are marked in the bottom axis.} 
\label{fig_sn_gpair}
\end{figure}

The first low-$T$ GDR measurement was done \cite{Baum98} in the case of $^{120}$Sn and the low-$T$ results \cite{Baum98,Heck43} for this nucleus have been used as a benchmark for several theories. Our region of interest in this work is the low-$T$ one where the earlier TSF calculations are not yielding satisfactory results \cite{aruepl}. As a first step, we include the pairing correlations in the TSF calculations within the BCS approach, but with free energies corresponding to the GCE approach \cite{MORET40B}. Without considering the PF, the pairing gaps averaged over the shapes could be obtained (\ref{average_1}). The results of TSF calculations done in this way are presented in Fig.~\ref{fig_largeG}, where the pairing calculations are carried out as mentioned in Sec.~\ref{p_constant1}
(marked with the legend $\tilde G$) and Sec.~\ref{pconstant_2} (marked with the legend $G$). The first interesting point is the development of proton pairing (for $Z=50$) even at low temperatures only when we use $\tilde G$. This is due to two reasons, namely (1) use of high pairing strength and (2) that pairing appears in deformed shapes, which can contribute when we average over the deformations. 

The former effect is mostly unphysical because, for
closed-shell nuclei, no scattering in to the empty orbitals in the next major shell is possible. This effect is also an artifact of the choice of a large
number of orbitals for pairing calculations (Sec.~\ref{p_constant1}). If we choose a restricted number of orbitals as discussed in Sec.~\ref{pconstant_2} then, independent of the choice of $G$, the pairing gap for a closed shell is always zero.

The latter reason is interesting because it shows that in some cases, although the equilibrium shape is spherical and that of a closed shell, due to the thermal shape fluctuations the deformed shapes with pairing may contribute to the averaged values. In Fig.~\ref{fig_largeG}(b) the average quadrupole deformations ($\langle \beta \rangle$) are presented as a function of $T$. We can notice that the presence of proton pairing at low $T$, in $^{120}$Sn, leads to a sharp increase in $\langle \beta \rangle$ with its value raising from $\sim 0.1$ to $\sim 0.2$.  The average deformation of $^{120}$Sn at low $T$ being $\sim 0.2$  is unrealistic, especially considering the fact that such a large deformation will yield a larger GDR width in contrast
to the experimental trend. Also we note that, with $\tilde G$, the pairing sets in, even for the deformation $\beta =0.1$.  At this deformation the variation of pairing gap as a function of pairing strength is depicted in Fig.~\ref{fig_sn_gpair}, where we have marked the values of the pairing strengths obtained from Strutinsky calculations ($\widetilde{G}$) and also the prescription of Ref.~\cite{Nilsson} ($G$). The values obtained by using $\widetilde{G}$
are quite larger in comparison to those given by other prescription. If we consider the argument that pairing should lower the GDR width at low-$T$ (as we know from the microscopic calculations for $^{120}$Sn, for example as in Ref.~\cite{Dang044303}) to explain the experimental value, then the choice with $\tilde G$ has to be ruled out.
Hence, for further calculations, we choose the pairing strengths from Ref.~\cite{Nilsson}. With this prescription we do not have proton pairing in the deformed shapes either, and hence the averaged pairing gap for protons is zero at all temperatures as shown in Fig.~\ref{fig_largeG}(a). The resulting $\langle \beta \rangle$ with pairing get attenuated only at very low temperatures where the neutron pairing still exists. This lowering of $\langle \beta \rangle$ indicates the lowering of GDR width as a consequence of the pairing effect. The existence of pairing at higher temperatures should decrease the width by those temperatures as well. Such nonvanishing or prolonged pairing gaps are predicted, e.g., in Refs.~\cite{Dang014318,MORET40B} by the calculations which include fluctuations in the pairing.

\begin{figure}
\includegraphics[width=0.55\columnwidth, bb=8 1 532 573]{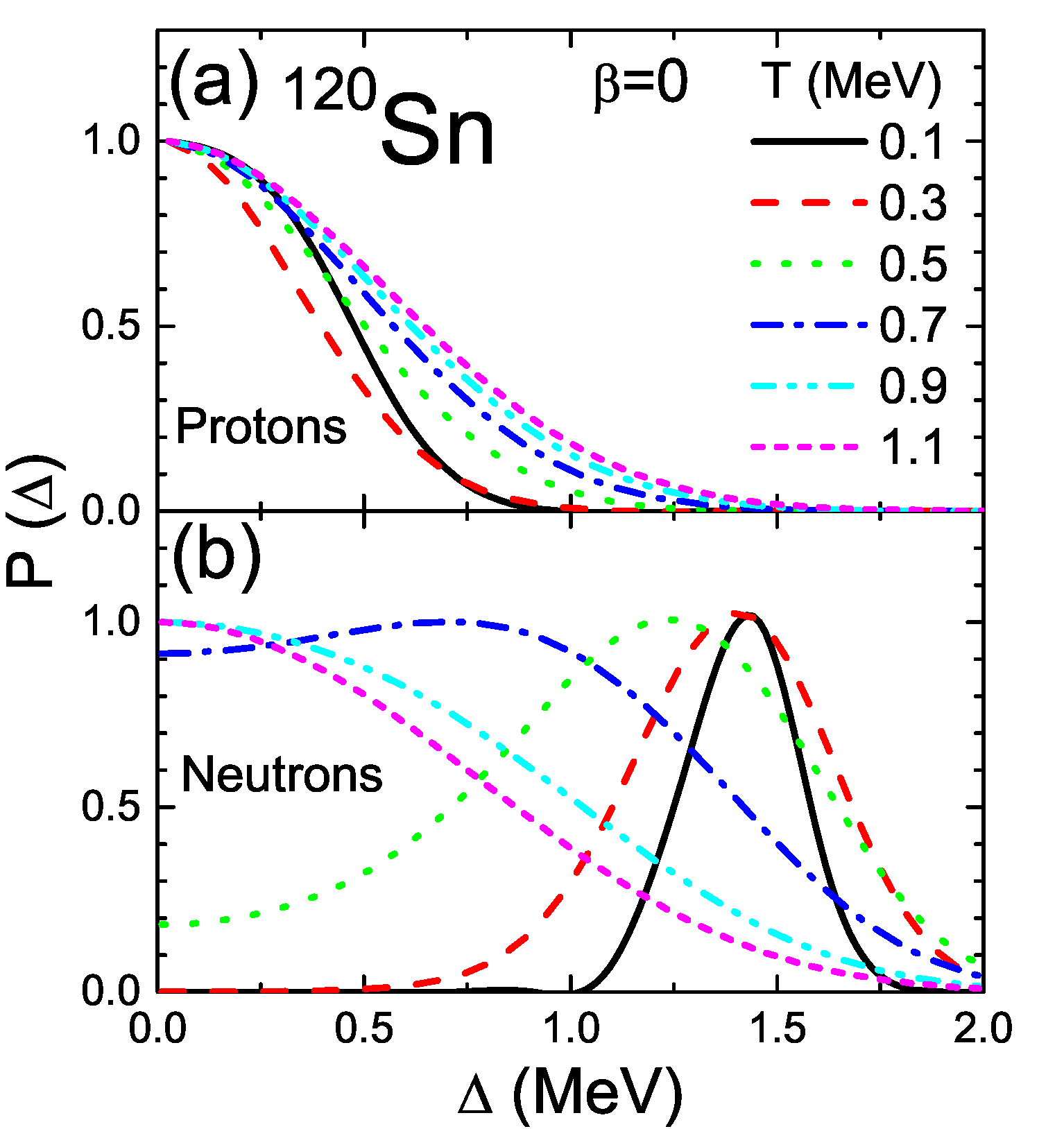}
\caption{(Color online) The probability distribution $P(\Delta
)$ for gap parameter $\Delta$ of protons and neutrons in the case of $^{120}$Sn with $\beta$ = 0 at different temperatures. $P(\Delta
)\varpropto \exp (-F_\mathrm{TOT}/T)$ where $F_\mathrm{TOT}$ corresponds to Eq.~(\ref{free energy}) and we have normalized the peak of $P(\Delta)$. The simple BCS calculations yielding $\Delta=0$ translates (in the calculations with pairing fluctuations) into the scenario where $\Delta=0$ is the most probable value and the occurrence of $\Delta>0$ has a small but finite probability.}
\label{fig_sn_dfluc}
\end{figure}

As mentioned in the theoretical framework, the PF could be accounted for by assuming a grand canonical partition function where the particle-number fluctuation is allowed. The pairing gaps are averaged using the  expression%
\begin{equation}
\langle \Delta_{i}\rangle _{\Delta _{P},\Delta _{N}}=\frac{\int_{\Delta_{P}}\int_{\Delta_{N}} \mathcal{D}%
[\alpha ]\exp \left[{-F_\mathrm{TOT}(T;\beta ,\gamma ,\Delta _{P},\Delta _{N})/T}\right]\Delta_{i}}{\int_{\Delta_{P}}\int_{\Delta_{N}}
\mathcal{D}[\alpha ]\exp \left[{-F_\mathrm{TOT}(T;\beta ,\gamma ,\Delta _{P},\Delta _{N})/T}\right]}\; (i= P, N),
\label{eq43}
\end{equation}%
with $\mathcal{D}[\alpha ]=\Delta _{P}\ \Delta _{N}\ d\Delta _{P}\ d\Delta _{N}$.

As mentioned in Ref.~\cite{MORET40B}, the gap parameter values obtained from the gap equation are to be understood as the most probable ones. The probability distribution for a given $\Delta$ can be calculated using $P(\Delta)\varpropto \exp (-F_\mathrm{TOT}/T)$ [where $F_\mathrm{TOT}$ corresponds to Eq.~(\ref{free energy})]. These probabilities at various $T$ for the case of protons and neutrons in $^{120}$Sn are plotted in Fig.~\ref{fig_sn_dfluc}. These results are quite consistent with previous results reported in Refs.~\cite{Dang014318,MORET}, justifying our pairing calculations with the grand canonical partition function. In the fluctuation calculations at very low $T$, due to the Boltzmann factor [$\exp(-F_\mathrm{TOT}/T)$], the most probable value (of either $\beta$ or $\gamma$ or $\Delta$) is favoured while we integrate over all possible values. Also, for the lowest value considered for $T$ (= $0.1$ MeV, below which we face numerical instabilities) one would expect that the fluctuations should not play a role and hence the averaged gap has to match with the most probable gap. At higher $T$, the Boltzmann factor widens and allows a strong contribution from the values closer to the most probable ones. 

\begin{figure}[tbp]
\includegraphics[width=.95\columnwidth]{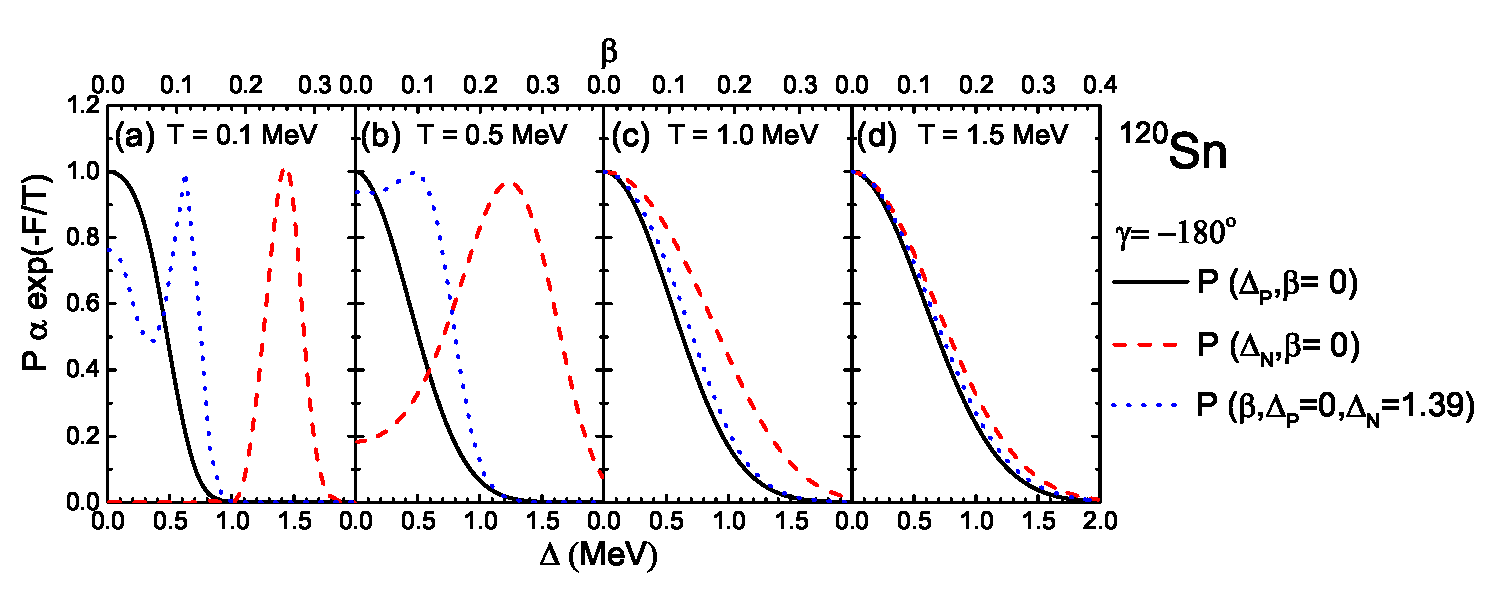}
\caption{(Color online) Variation of different probabilities ($P$) with respect
to the order parameters in the nucleus $^{120}$Sn for a fixed triaxiality parameter $\gamma= -180^\circ$. Solid and dashed lines represent the probability distributions for the gap parameter $\Delta$  of protons  and neutrons, respectively, where we fix $\beta$ = 0. Dotted lines represent the probability distributions for deformation parameter $\beta$ where the pairing gaps are fixed at their BCS values ($\Delta_P=0$ and $\Delta_N=1.39$). Different temperatures  are assumed in different panels as in the inset. $P(\Delta \text{ or } \beta)\varpropto \exp (-F_\mathrm{TOT}/T)$ where $F_\mathrm{TOT}$ corresponds to the Eq. (\ref{free energy}) and we have normalized the peak.  In the top
axis, the maximum value of $\beta$ is chosen such that 
$F_{\mathrm{TOT}}(\beta_{max})-F_{\mathrm{TOT}}(\beta=0)\approx 
F_{\mathrm{TOT}}(\Delta_P=2.0)-F_{\mathrm{TOT}}(\Delta_P=0.0)$.}
\label{fig_exp_del_beta}
\end{figure}

Having stated the importance of both PF and TSF, now we proceed to have a comparison between the strengths of these two fluctuations. In Fig.~\ref{fig_exp_del_beta}, we plot the probability distribution of pairing gaps with a fixed shape (at
$\beta=0$) along with the probability distribution of deformation parameter
$\beta$ with the pairing gaps frozen at their BCS values. The neutron and
proton pairing fluctuations are represented separately and these plots are
shown at different temperatures. To enable comparison between the changes
due to $\beta$ and $\Delta$, the upper limit of the $\beta$ values in
each panel are fixed by demanding equal energy difference with the change
in $\Delta_P$ values in the lower $x$ axis. The free energy is more sensitive
to the $\Delta_P$ than $\Delta_N$ because of the  nature of the proton closed
shell. We can see that both the PF and TSF are equally important at $T=0.1$
MeV. At $T=0.5$ MeV, the PF dominate mainly because of the neutron pairing. At $T=1.0$ MeV, the PF are relatively (when compared to previous case) less dominant. The PF and TSF again become equally important at $T=1.5$ MeV, where
the pairing field weakens and more excitation energy is available for the
nuclei to sample higher lying states of larger deformation. For the same
reason, at higher $T$ (not shown here), eventually the TSF become increasingly dominant as the PF saturate. Importantly we infer that it is necessary to consider both PF and TSF in low-$T$ calculations.    

\begin{figure}[tbp]
\includegraphics[width=0.55\columnwidth]{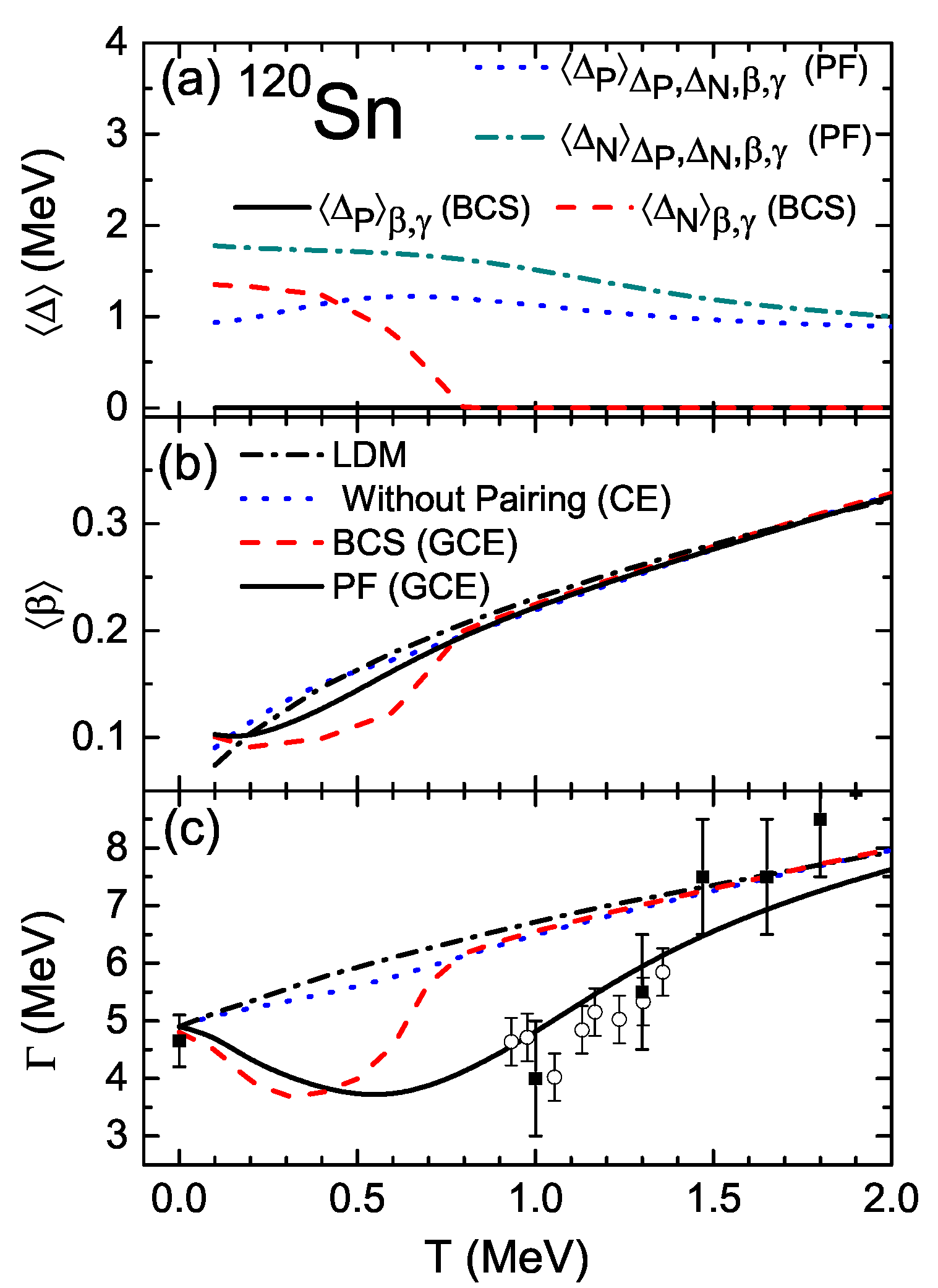}
\caption{(Color online) (a) Average pairing gap, (b) average quadrupole deformation parameter, and (c) GDR width in the case of $^{120}$Sn, as a function of temperature. The calculations  done without pairing utilize a CE approach (CE), whereas the calculations with  simple BCS pairing (BCS) and with pairing fluctuations (PF) utilize a GCE approach (GCE). The results obtained by using the liquid drop model (LDM) are also presented. Experimental values for $^{120}$Sn, taken from Refs.~\cite{Baum98,Heck43}, are shown by solid squares. For comparison, data for $^{119}$Sb, taken from Ref.~\cite{mukhu9}, are  also shown with open circles.}
\label{fig_snwidth}
\end{figure}

\begin{figure}[tbp]
\includegraphics[height=0.55\textwidth, angle=360]{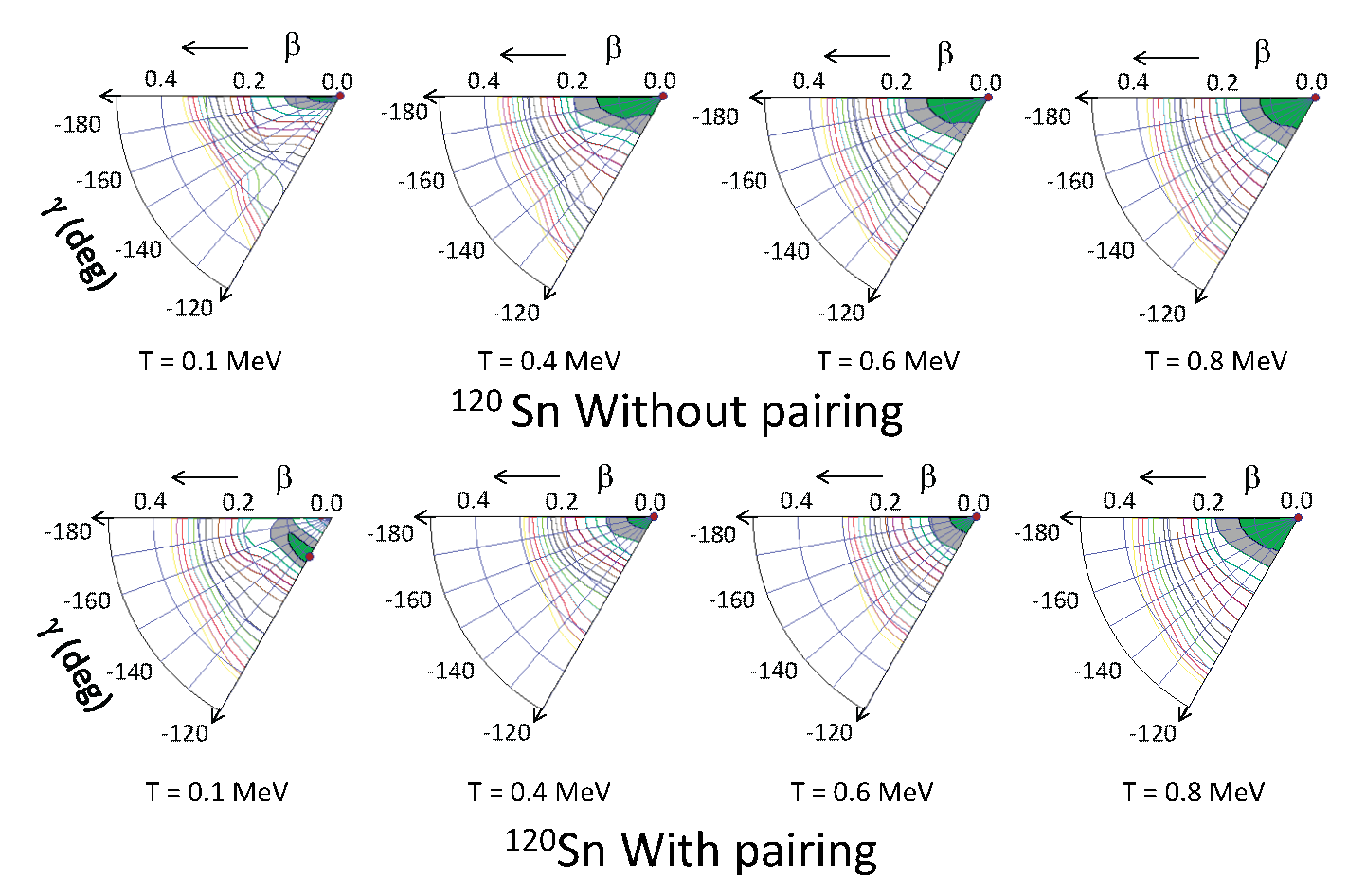}
\caption{(Color online) The free energy surfaces for $^{120}$Sn at different temperatures are plotted without pairing (CE) and with pairing (GCE). The contour line spacing is 0.5 MeV, the most probable shape is marked by a solid red circle and the first two minima are shaded.} 
\label{fig_sn_pes}
\end{figure}

Now it is interesting to see how these PF combine with the TSF. We carried out such calculations using free energies~(\ref{free energy}) from the grand partition function with averaging over the shapes as well as the pairing gaps for protons and neutrons. These results are presented in Fig.~\ref{fig_snwidth}, where we see that the combined effect of PF and TSF (shown with the legend PF) is quite different from the results of TSF with pairing included through the BCS approach (shown with the legend BCS). It has to be noted that now when we include pairing, we do the calculations within the GCE (\ref{free energy}), and when we neglect pairing, we resort to the CE (\ref{free energy_CE}). We can notice  from Fig.~\ref{fig_snwidth}(a) that, while having BCS pairing along with TSF we do not have the proton pairing gaps; but with the inclusion of PF the proton pairing develops and sustains even at $T=2$ MeV. In addition to the averaging over $\Delta$, now we have contribution from deformed shapes also due to the additional averaging over shapes (TSF). The single-particle energies change with deformation and at higher deformations we may not have a closed shell any longer. This may lead to a situation that the role of pairing is now enhanced even for a closed shell when we consider TSF in combination with PF. This enhancement is clearly seen when we compare the $\langle\Delta_P\rangle$ in Fig.~\ref{fig_snwidth}(a). 

From Figs.~\ref{fig_snwidth}(b) and~\ref{fig_snwidth}(c) it can be seen that both $\langle \beta \rangle$ and $\Gamma$ are quenched with the introduction of pairing but with PF, $\Gamma$ is more quenched than $\langle \beta \rangle$.
It has to be noted that the role of pairing on $\langle \beta \rangle$ is only through the free energy surface whereas $\Gamma$ is also affected by the term $\chi {\mathcal{P}}^\dagger {\mathcal{P}}$ [Eq.~(\ref{gdr_frequency})]. This additional term along with prolonged pairing in the case of PF leads to the quenching of $\Gamma$ at higher temperatures when compared to the case of BCS.  Thus, with the inclusion of PF, we can explain the experimental results very well. The calculations without pairing and with BCS could not explain the data for $\Gamma$ at $T=1.0$ and 1.3 MeV. In the GDR calculations with PF, for the resulting larger pairing gaps, the parameter $\chi$ has to be readjusted. Instead, in a more effective way, we retain the same value for $\chi$ but subtract the asymptotic (high $T$) value of $\Delta$ (typically $\sim 0.75$ MeV) from the calculated $\Delta$ entering Eq.~(\ref{omega_P}). 
 
The prolonged pairing in the PF case plays no role in the free energy surfaces and hence $\langle \beta \rangle$ obtained by using PF reaches the value corresponding to the one without pairing at the $T$ where the BCS pairing gap vanishes (i.e., where it stops contributing to the free energy). In Fig.~\ref{fig_snwidth}(b) we can notice that the suppression of $\langle \beta \rangle$ in the BCS case is more than that given in the case with PF. This situation can be understood by examining the corresponding free energy surfaces presented in Fig.~\ref{fig_sn_pes}.  These surfaces show minima at zero deformation while ignoring the pairing at all temperatures, and we get a nonzero $\langle \beta \rangle$ due to the TSF. As $T$ increases the TSF gains more strength leading to a larger
$\langle \beta \rangle$, even if there is no change in the free energies. In the present case, while $T$ increases, the first two minima in the free
energy surfaces span a larger area and hence compound the increase in fluctuations resulting in the increase of $\langle \beta \rangle$. At $T=0.1$ MeV, the inclusion of pairing shifts the most probable deformation to a weakly deformed one from a spherical case.  This leads to a small increase in $\langle \beta\rangle$ at this point. At $T=0.4$ and 0.6 MeV, we can see that the effect of pairing is to make the minimum crisp (or deep) which inhibits the shape fluctuations.  Therefore, $\langle \beta \rangle$ is suppressed at these temperatures, and at $T=0.8$ MeV where the BCS pairing gap vanishes, the free energy surfaces with and without pairing look the same. If we include PF in addition to the
role of BCS pairing, the TSF is enhanced and hence leads to a higher value
of $\langle\beta \rangle$ when compared to the TSF with BCS pairing gaps.
We can also note that the free energies without and with pairing are calculated with CE and GCE, respectively. When there is no pairing, the relative energies, as depicted in Fig.~\ref{fig_sn_pes} ($T=0.8$ MeV), do not clearly distinguish
between these two ensembles.

\subsection{$^{179}$Au}
\begin{figure}[tbp]
\includegraphics[width=0.55\columnwidth]{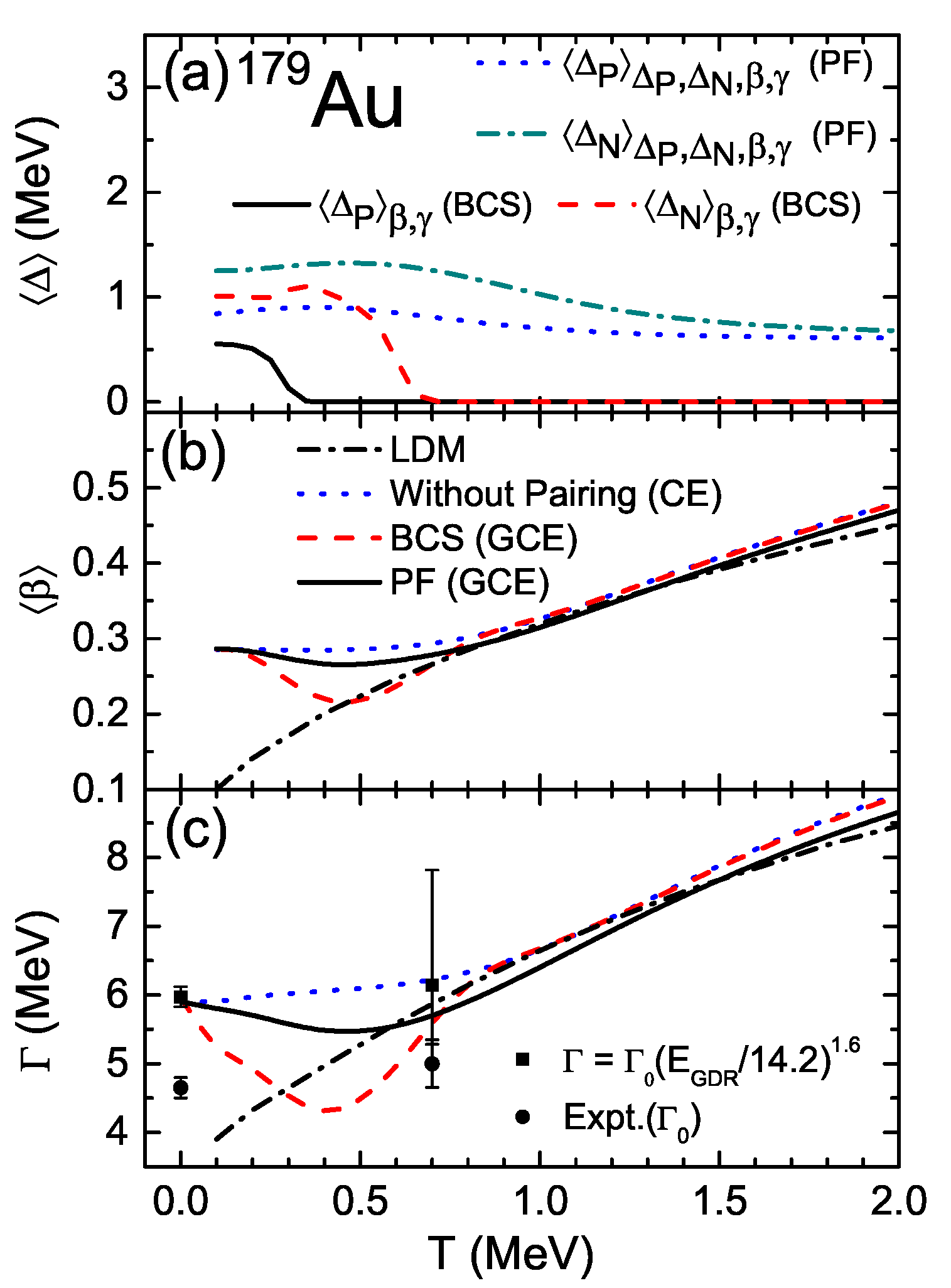}
\caption{(Color online) Similar to Fig.~\ref{fig_snwidth} but for the nucleus $^{179}$Au. Experimental values for the spherical width ($\Gamma_{0}$)  are taken from Ref.~\cite{camera155} and are shown with solid circles. The corresponding deformed widths ($\Gamma$) are estimated using the relation $\Gamma=\Gamma_{0}(E_{\mathrm{GDR}}/14.2)^{1.6}$ and are shown with solid squares.}
\label{fig_auwidth}
\end{figure}

\begin{figure}[tbp]
\includegraphics[height=0.55\textwidth, angle=360]{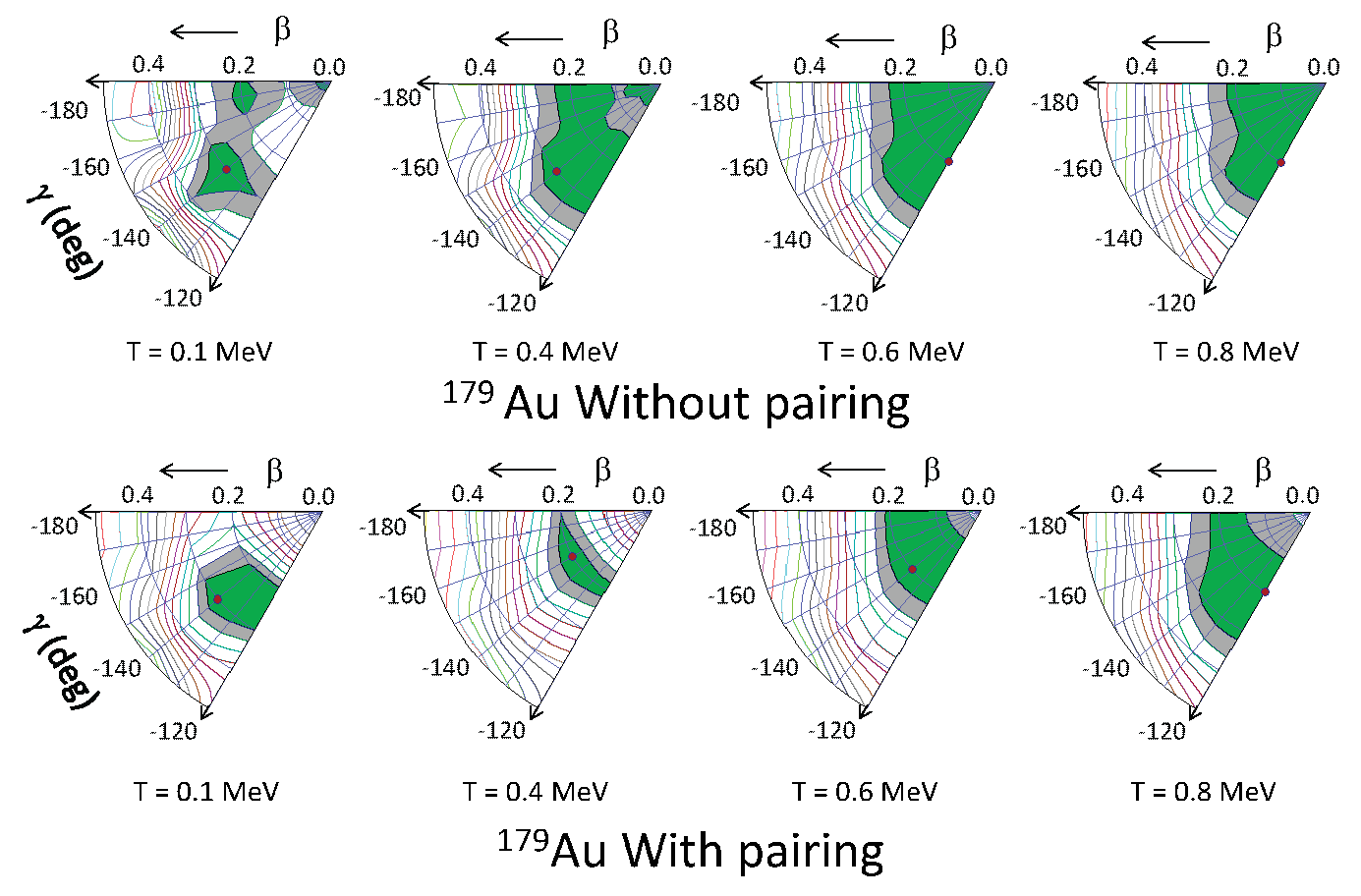}
\caption{(Color online) Similar to Fig.~\ref{fig_sn_pes} but for the nucleus $^{179}$Au.} \label{fig_auallpes}
\end{figure}

Having analyzed the role of pairing and its fluctuation on GDR properties
in a spherical nucleus, we proceed to study the deformed nucleus $^{179}$Au. TSF calculations (without pairing) suggested a strong enhancement
of GDR width at low $T$ \cite{aruepl} owing to the deformation effects, whereas the experimental observation \cite{camera155} pointed otherwise. Here we explore how the inclusion of pairing and PF could change the previous interpretations.  The calculated averaged  pairing gaps in different cases for the nucleus $^{179}$Au, are shown in Fig.~\ref{fig_auwidth}(a) as a function of $T$. We can see that  pairing plays a strong role in this nucleus as depicted by the large averaged pairing gaps for both protons
and neutrons. The simple BCS treatment suggests that the proton and neutron pairing gaps vanish at $T\sim0.4$  and 0.7 MeV, respectively.  While considering the PF along with TSF, the  averaged  pairing gaps continue to be strong even at $T=2$ MeV, similar to the case of $^{120}$Sn.

 The $\langle\beta \rangle$ and $\Gamma$ get quenched at low $T$ as seen
in Figs.~\ref{fig_auwidth}(b) and~\ref{fig_auwidth}(c), respectively.  Similar to the  case of $^{120}$Sn, the decrease in $\langle\beta \rangle$  with BCS calculations can be understood from the corresponding free energy surfaces which are shown in Fig.~\ref{fig_auallpes} for different
temperatures and calculated without and with pairing. At $T=0.1$ MeV, the calculations without pairing suggest coexisting shapes
corresponding to spherical, oblate, and triaxial  shapes. For such a situation allowing variety of shapes (shallow minimum or multiple minima in the free energy surface) to contribute, the TSF leads to an increase in the averaged values ($\langle\beta \rangle$ and $\Gamma$). 
The inclusion of pairing in the deformation energies yields a deeper minimum corresponding to a  triaxial shape. At higher    temperatures  also, this trend continues to exist. From these free energy surfaces, we can say that
with the inclusion of pairing $^{179}$Au can have only a narrow range of deformation, which means that there will be a quenching in the $\langle\beta \rangle$ (which will be carried forward to $\Gamma$) when compared to calculations without pairing. At $T=0.4$ MeV the narrowing of the minimum is strong and hence shows a strong suppression in $\langle\beta \rangle$. At this $T$, the proton pairing vanishes and marks the beginning of the upward trend for $\langle\beta \rangle$. In case of $T=0.6$ and $0.8$ MeV, without pairing, our calculations suggest a shallow bottom in the free energy surface spanning a wide range of deformations with the minimum at a large deformation with $\beta=0.2$ and $\gamma=-120^\circ$.  The corresponding calculations with pairing also suggest a similar shape (with marginal preference to a deformed shape) because of weaker contribution from the pairing to the deformation energies. The $\langle\beta \rangle$ from PF calculations are largerr than that of BCS calculations, for the same reason as discussed in the case of $^{120}$Sn.

All the effects discussed above, governing the variation of $\langle\beta \rangle$, are carried forward in our results for $\Gamma$ as shown in Figs.~\ref{fig_auwidth}(b) and ~\ref{fig_auwidth}(c). However, for $\Gamma$, there is an additional pairing effect in the GDR Hamiltonian. It is convenient to consider the following factors which can affect the $\Gamma$ in our calculations \cite{Rhineprc}
\begin{enumerate}
\item The pairing effects:
\begin{enumerate}
\item modification of free energy surfaces;
\item damping GDR frequencies through the term $\chi {\mathcal{P}}^\dagger {\mathcal{P}}~(\ref{GDRhamiltonian}).$
\end{enumerate}
\item The shell effects (in comparison to liquid drop behavior).
\end{enumerate}
The free energy surfaces of $^{179}$Au and $^{120}$Sn show the same trend
while including the pairing effect. By including pairing, the
minimum in the free energy surfaces becomes deeper and hence the TSF are attenuated.
 This results in the suppression of $\langle\beta \rangle$ and $\Gamma$. Similarly  factor 1(b) listed above also has the same role (suppressing $\Gamma$) in both $^{179}$Au and  $^{120}$Sn.  Factor 2 is expected to decrease the $\Gamma$ in $^{120}$Sn by favoring spherical shape, but this
effect is almost negligible as we see that the results of LDM and CE are
almost same in Fig. \ref{fig_snwidth}(c). However, we observe that this factor increases $\Gamma$ in $^{179}$Au. Due to this, the overall suppression of $\Gamma$ (at lower $T$, in comparison with $T=0$) is smaller in $^{179}$Au when compared to that of $^{120}$Sn.

In Fig.~\ref{fig_auwidth}(c), we have also plotted the experimental values
\cite{camera155}, which are ambiguous. In Ref.~\cite{camera155}, the  (spherical)
GDR width $\Gamma_{0}$ of  $^{179}$Au at $T = 0.7$ MeV is quoted as $5.0\pm0.35$ MeV  with a deformation parameter $\beta= 0.1\pm0.1$ and the (spherical) GDR\ centroid energy $E_0=14.2$ MeV.  It may not be appropriate to compare this $\Gamma_{0}$ with the (deformed) GDR width ($\Gamma$) obtained theoretically. For a better comparison, we have estimated the equivalent $\Gamma$ using the relation \cite{camera155} $\Gamma=\Gamma_{0}(E_{\mathrm{GDR}}/E_0)^{1.6}$ along with the experimentally suggested values for $\Gamma_{0}$,  $E_0$, and  $\beta= 0.1$. In a similar way we have estimated the  width at $T = 0$ MeV  from a given $\Gamma_{0}$ ($4.65\pm0.15$ MeV)~\cite{camera155}, with $\beta= -0.139$~\cite{moller2}. This conversion results in large error bars which encompass all the results corresponding to different calculations (without pairing, BCS, and PF). The Goldhaber-Teller model calculation~\cite{RIPL3Au179} with empirical inputs, yields a larger width ($\sim$5.7 MeV) compared to the experimental $\Gamma_{0}$ of $5.0\pm0.35$ MeV. All these facts raise the ambiguity while corroborating the results of theory and experiment. We also notice that, unlike the case of $^{120}$Sn, the low-$T$ data for $^{179}$Au are not sufficient to have a constraint on the parameter $\chi$ or to conclude whether a pairing approach is necessary or not. Thus, there is indeed a need for more experimental data on GDR width at low temperatures. 

\subsection{$^{208}$Pb} 

\begin{figure}[tbp]
\includegraphics[width=0.55\columnwidth]{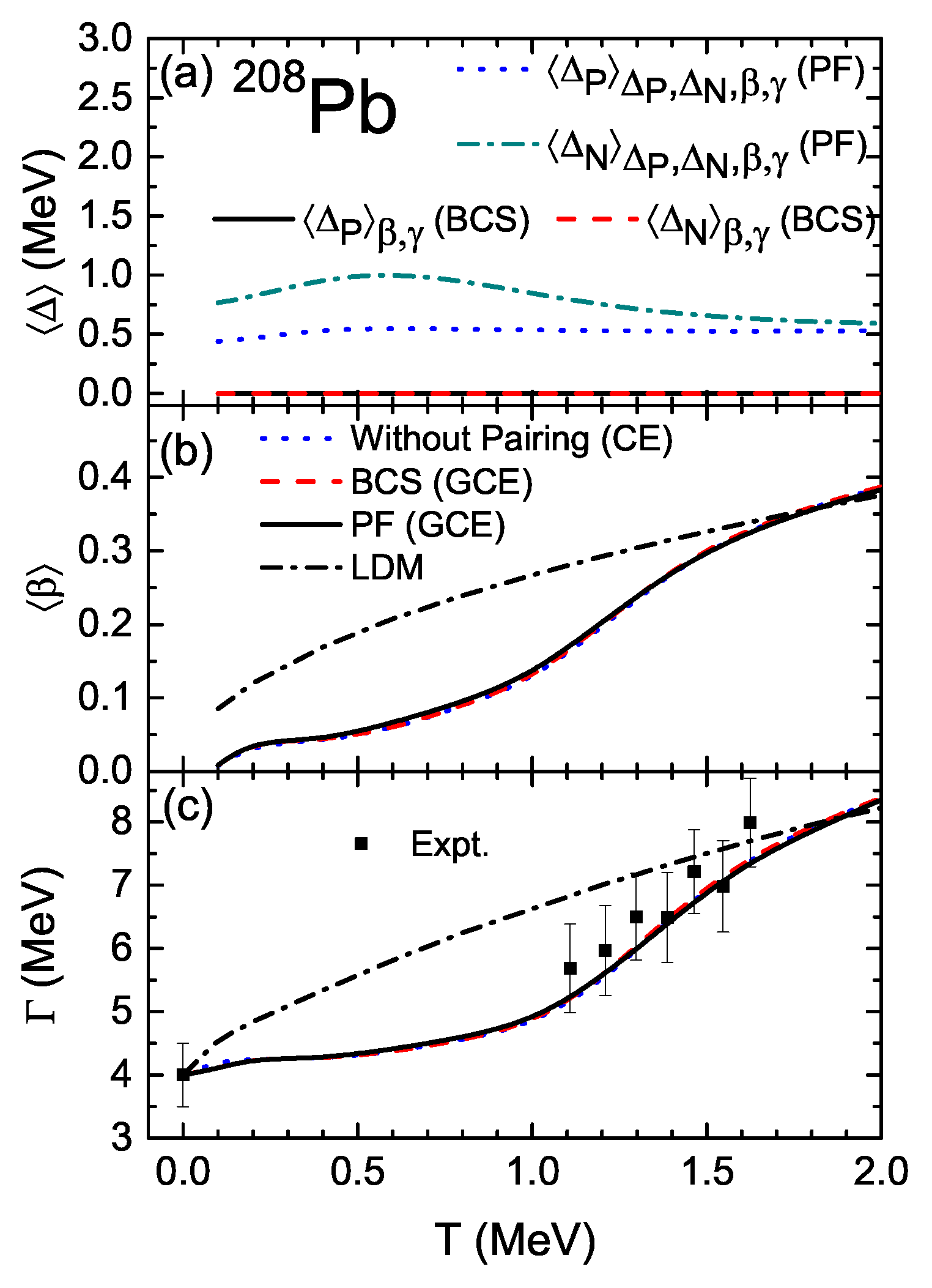}
\caption{(Color online) Similar to Fig.~\ref{fig_snwidth} but for the nucleus $^{208}$Pb. Experimental values shown with solid squares correspond to those given in Refs.~\cite{Kusn98,Baum98}.}
\label{fig_Pbwidth}
\end{figure}

The PF are observed to be important even in the spherical nucleus $^{120}$Sn where we have a closed shell for protons. In order to see the role of PF in  a doubly closed shell nucleus, we carried out the calculations for the nucleus $^{208}$Pb and the results are shown in Fig.~\ref{fig_Pbwidth}. As expected for closed shells, the most probable values remain zero for pairing gaps for both protons and neutrons. With the inclusion of PF we obtain finite $\langle\Delta\rangle$  for both the protons and neutrons and they last up to $T=2$ MeV. The marginal increase in  $\langle\Delta\rangle$ at lower $T$ is due  to  reasons similar to those discussed in the case of protons in $^{120}$Sn. The $\langle\beta\rangle$ estimated with LDM shows a larger value when compared with other calculations which include the shell corrections. The PF are found to have negligible contribution to the $\langle\beta\rangle$
despite appreciable $\langle\Delta_N\rangle$ at lower $T$.  At $T=0.5$ MeV,
\ $\langle\Delta_N\rangle\sim 1.0$ MeV and still there is not appreciable
effect of PF.  For closed shells, the free energies are stiff with respect
to $\Delta$ and hence the corresponding Boltzmann factor will be sharp, leading
to a suppression of PF. All the effects observed in $\langle\beta\rangle$
are reflected in $\Gamma$. Thus, in $^{208}$Pb, the PF has no role on $\Gamma$
and the shell effects strongly quench $\Gamma$ at low $T$.
This observation is in good accordance with experimental data available for
$T\gtrsim 1.1$ MeV. Measurements at lower $T$ and with more accuracy can strengthen the arguments regarding the role of PF and shell effects.

\subsection{$^{97}$Tc}
\begin{figure}
\includegraphics[width=0.55\textwidth, angle=360]{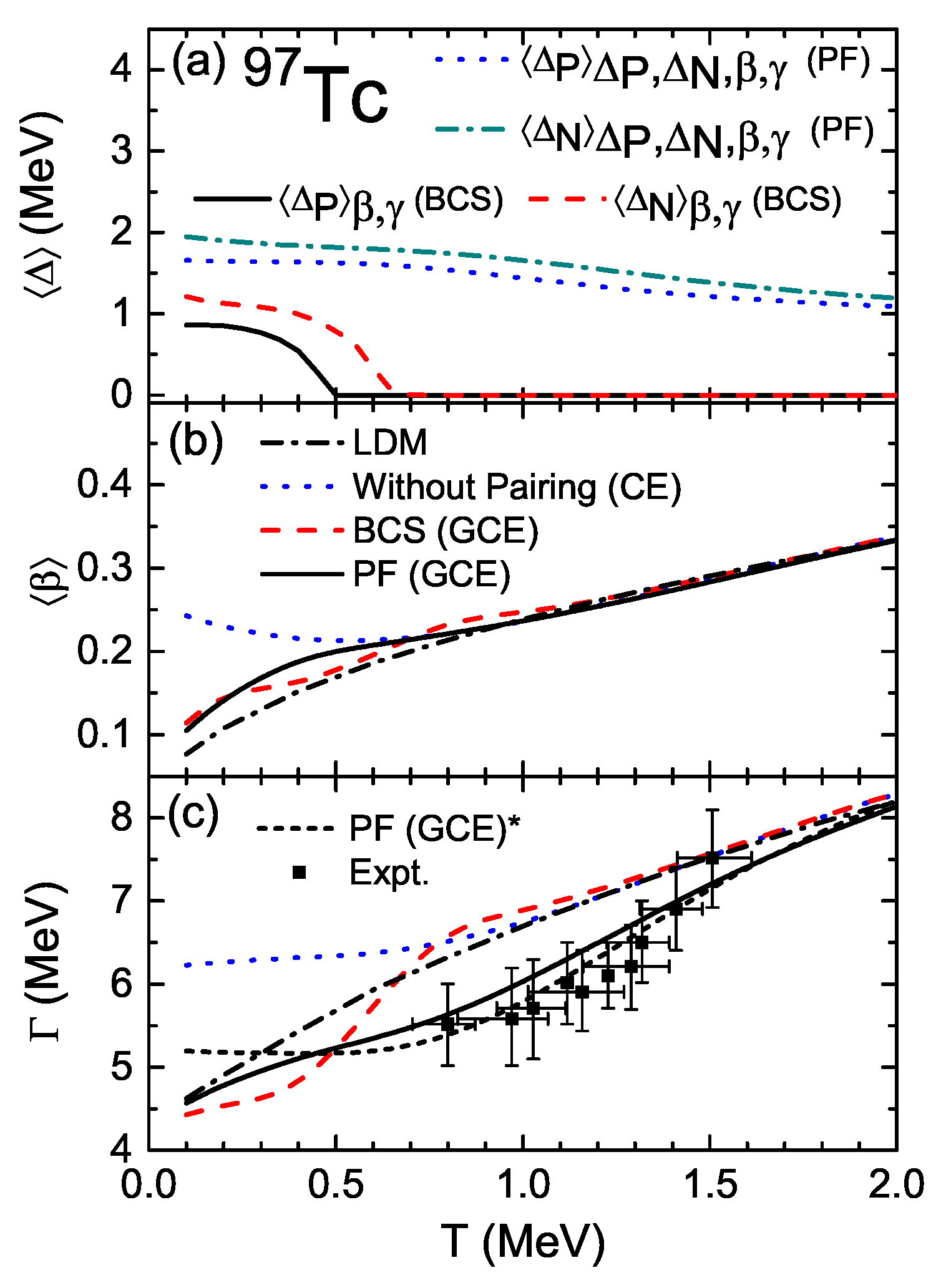}
\caption{(Color online) Similar to Fig.~\ref{fig_snwidth} but for the nucleus $^{97}$Tc. Experimental values shown with solid squares correspond to those given in Ref.~\cite{Balaram}. The legend PF (GCE)* denotes the results obtained
in the calculations by using the parameter $\delta=1.9$. In all other calculations $\delta=1.8$.} \label{fig_Tc97width}
\end{figure}

\begin{figure}
\includegraphics[height=0.55\textwidth, angle=360]{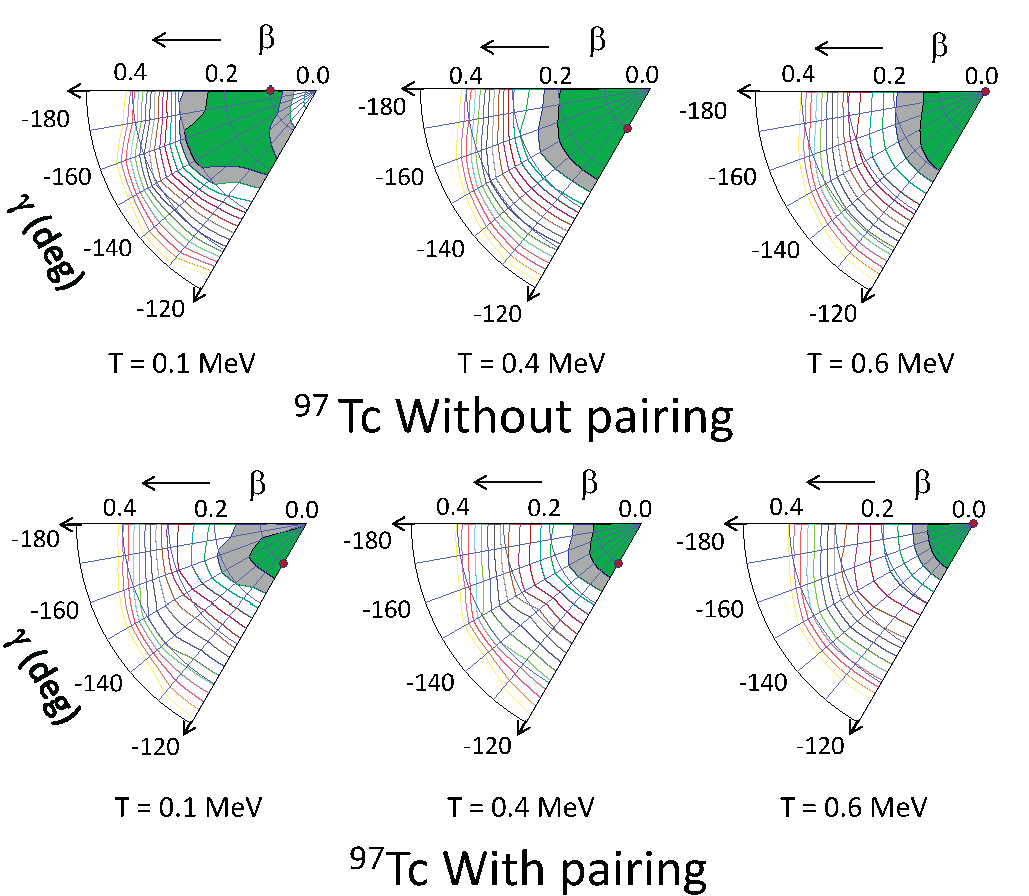}
\caption{(Color online) Similar to Fig.~\ref{fig_sn_pes} but for the nucleus $^{97}$Tc.} \label{fig_Tcpes}
\end{figure} 

\begin{figure}[tbp]
\includegraphics[height=0.55\textwidth, angle=360]{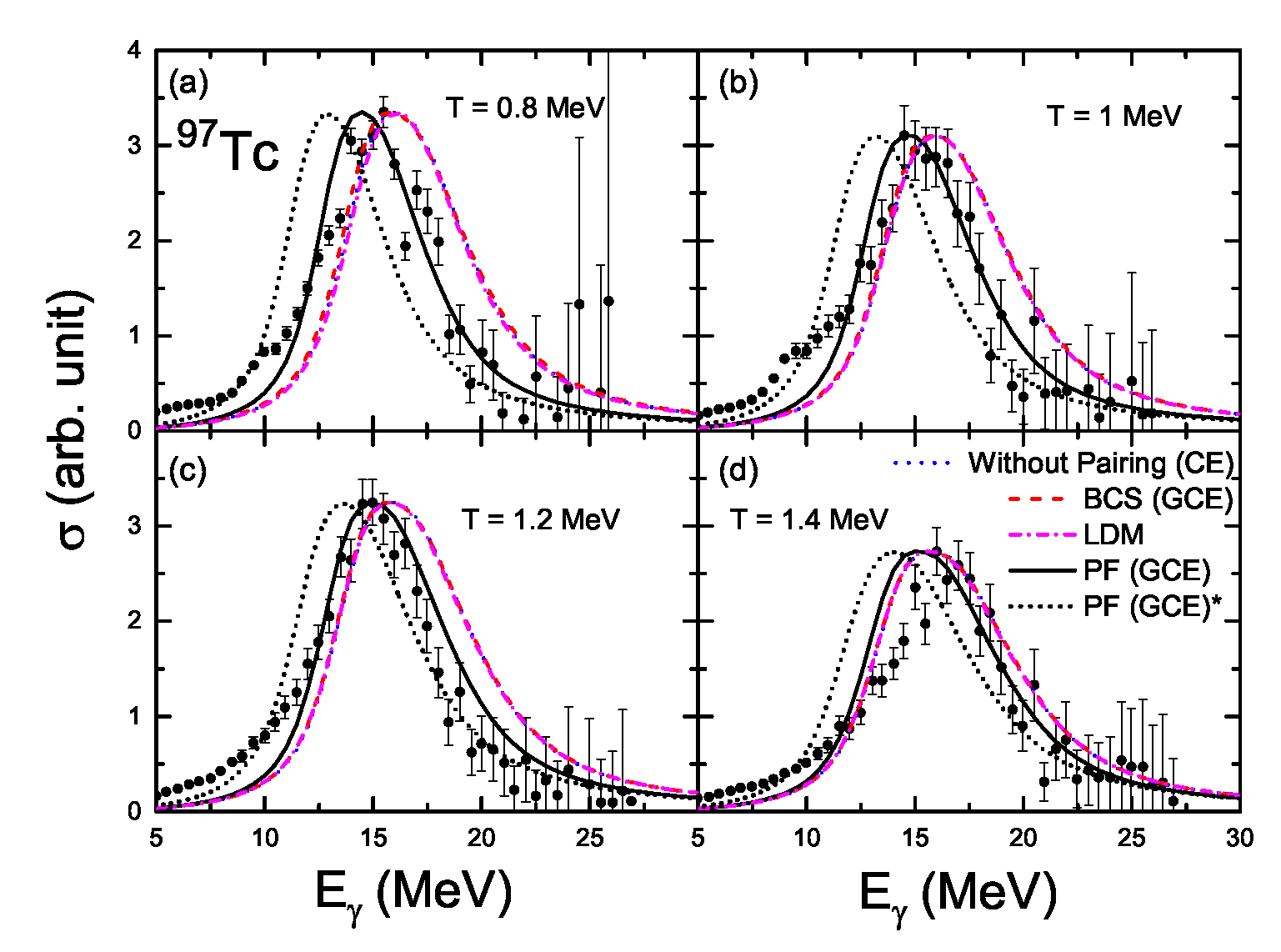}
\caption{(Color online) The GDR strength functions $^{97}$Tc at different temperatures ($T$) are compared with the results obtained using the pairing fluctuations (PF) within a grand canonical ensemble approach (GCE).  Experimental data are taken from  Ref.~\cite{Balaram} and are shown by solid circles. The legend PF (GCE)* denotes that the calculations are with the parameter $\delta=1.9$; in all other calculations $\delta=1.8$.} 
\label{fig_Tc_cross}
\end{figure} 

Very recently, experimental data at low $T$ for the nucleus $^{97}$Tc~\cite{Balaram} are reported and some of our results for this  nucleus are presented in \cite{Rhineprc}. Here we discuss our results elaborately with the aid of free energy surfaces
and strength functions.  In $^{97}$Tc, the BCS calculations without fluctuations suggest that  the proton and neutron pairing gaps vanish at $T=0.5$ MeV and $T\sim0.65$ MeV respectively, as shown in Fig.~\ref{fig_Tc97width}(a). With both protons and neutrons in the mid shells, we obtain very large
values for $\langle\Delta\rangle$ from the calculations with the combination
of PF and TSF. The $\langle\Delta\rangle$ continues to be strong even at $T=2$ MeV. The $\langle\beta \rangle$ and $\Gamma$ get strongly quenched with the inclusion of pairing as shown in Figs.~\ref{fig_Tc97width}(b) and~\ref{fig_Tc97width}(c). This quenching can be understood from the free energy surfaces shown in Fig.~\ref{fig_Tcpes}. The calculations without pairing show a shallow minimum in the free energy surface, which allows a variety of shapes and enhances TSF. The inclusion of pairing strongly reduces the probable shapes as the area of the minimum gets narrower in the ($\beta, \gamma$) plane. Subsequently, the TSF with pairing leads to reduced $\langle\beta \rangle$ which is carried forward to $\Gamma$. The LDM free energies favor a spherical shape and hence yield a lower $\langle\beta \rangle$ when compared with the results without pairing (which have only shell effects as additional contribution). Thus the shell effects tend to increase the $\langle\beta \rangle$, but the pairing decreases $\langle\beta \rangle$ at lower $T$. The role of pairing in the free energies is only up to $T\sim0.65$ MeV, while including pairing without PF  leads to  quenching in $\langle\beta \rangle$ and $\Gamma$. Hence, similar to the case of $^{120}$Sn, only with the consideration of PF can we explain the experimental data. It is interesting to note that $\langle\beta \rangle_{\mathrm{LDM}}<\langle\beta \rangle_{\mathrm{PF}}$ at low $T$ whereas $\Gamma_{\mathrm{LDM}}>\Gamma_{\mathrm{PF}}$.
This is due to the fact that the role of pairing on $\Gamma$ through the
term $\chi\mathcal{P}^\dagger\mathcal{P}$ is very dominant in relation to
the role through $\langle\beta \rangle$. 

In Fig.~\ref{fig_Tc97width}(c) we have shown with a  short-dashed line the results from our calculation, where the parameter $\delta$ is chosen to be $1.9$. With such a choice, the $\Gamma$ lowers appreciably leading to an apparent improvement in the fit. However, due to the uncertainties in the data, this distinction is not clear. For a better and detailed corroboration with experimental data \cite{Balaram}, we calculated the GDR cross sections. The results for
selected representative experimental data are shown in Fig.~\ref{fig_Tc_cross}, where the theoretical results with PF show a fair agreement with the experimental data. All the other calculations without PF lead to a significant shift in the centroid energy and hence poorly agree with the data.  At $T=1.4$ MeV,  there is an apparent increase in the experimental GDR centroid energy.
This feature is inexplicable by our calculations but this discrepancy cannot
be weighed due to large uncertainties in the data around the peak.\ However,
if such a shift in centroid energy could be established precisely, it would
shed
more light on the role of pairing on the GDR centroid energy. The present uncertainties in the measured strength functions render the corresponding extracted width to be less reliable. For example, in all the cases shown in Fig.~\ref{fig_Tc_cross}, even if we have increased the width of the theoretical cross sections, the overall agreement with data could be of the same quality.  The calculations with $\delta=1.9$ leads to  cross sections
which  strongly deviate from the measured ones. Though  the data
for the $\Gamma$ could not be used to validate the choice of $\delta$, the
data for cross sections help us to fix the parameter $\delta$ distinctly. With the unambiguously determined parameters, the TSFM with the inclusion of PF can explain the GDR width as well as the GDR strength functions fairly
well at low temperatures. More precise data at low $T$ can be more informative
in this regard.

It is worth mentioning that, as has been pointed out by Gervais \textit{et al}.~\cite{Gervais}, the strong asymmetry of the GDR cross section predicted by the TSFM leads to a depletion of the strength in the low-energy region $E_\gamma< 12$ MeV. This feature is clearly seen in Fig.~\ref{fig_Tc_cross}. Including PF in this work improves the description of the GDR width at low
$T$, but cannot resolve the deficiency of the TSFM with respect to the description of the GDR line shape, because this deficiency in the current framework of the TSFM is even stronger at larger excitation energies where the effect of PF is negligible or completely gone.

So far, the TSFM has always followed a macroscopic approach to the
GDR where the components of the GDR cross sections are assumed to have an asymmetric form. For example, in our approach it is given by Eq.~(\ref{Eq.Cross}) and by a similar equation in Refs.~\cite{ALHAT,CARL_Sm}. An addition of such asymmetric components with an energy-dependent width is expected to yield an asymmetric GDR cross section, even for spherical nuclei
for which the experiments suggest a symmetric curve. Part of this discrepancy can be removed by choosing symmetric shapes for the GDR components, making use of the Breit-Wigner distributions, for  example.
However,  all the parameters in our approach have to be readjusted along with revisiting the energy dependence of the width.
Alternatively, combining a microscopic treatment of the GDR with the TSFM could resolve this issue. A more careful analysis in this regard is in progress.  
\begin{table}
\begin{tabular}{|c|c|c|c|c|c|c|}
\hline
\multirow{2}{*}{Nucleus}&(i) Shell & \multicolumn{3}{c|}{(ii) Pairing effects through $F_{\mathrm{TOT}}$}& (iii)Pairing effects & Net \\
\cline{3-5}
 & effects & $T=0.1$ MeV & $T=0.4$ MeV & $T=0.6$ MeV & through $\chi\mathcal{P}^\dagger\mathcal{P}$
 & effect \\  \hline

$^{120}$Sn&---        &$\uparrow$  & $\downarrow$&$\downarrow$&$\downarrow$&$\downarrow$\\

$^{179}$Au&$\uparrow$   &$\downarrow$  & $\downarrow$  &$\downarrow$&$\downarrow$&$\downarrow$\\

$^{208}$Pb&$\downarrow$ &---&--- &---&$\downarrow$&$\downarrow$\\

$^{97}$Tc &$\uparrow$   &$\downarrow$  & $\downarrow$  &---  &$\downarrow$&$\downarrow$\\

\hline
\end{tabular}
\caption{Change in GDR width ($\Gamma$) at typically low temperatures ($T\lesssim1$
MeV) due to (i) the shell effects, (ii)
pairing effects through the free energy ($F_{\mathrm{TOT}}$) at different $T$, and (iii) pairing effects through the attenuation of GDR frequencies in our macroscopic model for GDR [through the term $\chi\mathcal{P}^\dagger\mathcal{P}$ (\ref{GDRhamiltonian})].  The symbols $\uparrow$ and $\downarrow$ represent increase and decrease in $\Gamma$, respectively. This change for the effect (i) is with respect to a liquid drop model; for effects (ii), (iii) and
the net effect the change is with respect to the calculations without pairing
but inclusive of shell effects. The symbol --- represents no substantial change in $\Gamma$.}
\label{table2}
\end{table}

\section{Summary and Conclusion}
We have presented, in detail, our formalism to study giant dipole resonance (GDR) in nuclei at low temperature ($T$). This formalism is an extension of the thermal shape fluctuation model (TSFM) \cite{aruprc1,ALHA,ALHAT} with the proper treatment of pairing and its fluctuations. In a macroscopic approach, the GDR frequencies are related to the geometry (shape) of the resonating system as in a general cas,e but additionally we consider the role of pairing.  To compare with the measured GDR width, we consider the thermal fluctuations over the possible degrees of freedom, namely the shape and the pairing field. This is achieved through a weighted average of the observable over the considered degrees of freedom. The weights are given by the Boltzmann factor [$\exp(-F_{\mathrm{TOT}}/T)$] where the free energy ($F_{\mathrm{TOT}}$) is calculated within the Nilsson-Strutinsky (microscopic-macroscopic) approach where the $T$ dependence of the shell and pairing effects are taken care of properly. For calculations without pairing, we consider the canonical ensemble (CE) approach.

While considering pairing, to calculate $F_{\mathrm{TOT}}$, we consider the grand canonical ensemble (GCE)  for which the particle number fluctuations are inherent. In a simple BCS approach, the paring gap  vanishes at a critical $T$ (as in a second-order phase transition) but when we consider the pairing fluctuations (PF), where the averaging over pairing gap is also included, the average pairing gap smoothly varies and remains strong even at $T=2$ MeV. Such an extended superfluid phase is more pronounced while we treat PF in combination with shape fluctuations. Pairing not only contributes in modifying $F_{\mathrm{TOT}}$ but also plays a role in attenuating the GDR frequencies (and hence the width $\Gamma$) in our macroscopic model for GDR [through the term $\chi\mathcal{P}^\dagger\mathcal{P}$ (\ref{GDRhamiltonian})]. Hence, the pronounced and sustained pairing suggested by the calculations including
PF leads to a quenching of GDR width in agreement with the experimental observations. This  indicates that for small systems like the atomic nuclei, there are  no sharp superfluid-normal phase transitions \cite{MORET40B,Dang064315}. 

For a precise match with the $^{120}$Sn and $^{97}$Tc data, the consideration of PF is crucial, whereas the present $^{179}$Au data are not precise enough to arrive at a similar conclusion. In the nucleus $^{208}$Pb, the shell effects are so dominating as to favor a spherical (closed-shell/unpaired) configuration and hence the role of PF is negligible. The factors affecting $\Gamma$, namely (i) shell effects, (ii) pairing effects through $F_{\mathrm{TOT}}$, and (iii) pairing effects through $\chi\mathcal{P}^\dagger\mathcal{P}$ are summarized in Table \ref{table2} for all the nuclei considered in this work. It is interesting to note that, although the factor (iii) is monotonic, there is a strong interplay between factors (i) and (ii) which can vary from nucleus to nucleus. More studies are required to pin down the possible correlations.

We observe that the estimation of pairing force strength ($G$) in a Strutinsky way, by assuming an average pairing gap of $12/\sqrt A$, leads to an overestimation as the corresponding results are not consistent with the observed GDR data. In the case of $^{120}$Sn we have demonstrated this fact which opens up an idea of considering the precise low-$T$ GDR measurements to provide a benchmark for pairing prescriptions. 

\section*{Acknowledgments}
This work is supported by the Science and Engineering Research Board (India), SR/FTP/PS-086/2011. The financial support from the Ministry of Human Resource Development, Government of India, to one of the authors (A.K.R.K) is gratefully acknowledged. A part of this work was completed at RIKEN and the numerical calculations were carried out using RIKEN Integrated Cluster of Clusters
(RICC).

\end{document}